\documentclass[preprintnumbers,amsmath,amssymb,prd, notitlepage,nofootinbib,twocolumn, superscriptaddress]{revtex4}
\usepackage{aas_macros}
\usepackage{graphicx}
\usepackage{multirow}
\usepackage{amsbsy}
\usepackage{mathrsfs}
\usepackage{amsmath}
\usepackage[normalem]{ulem}
\usepackage{array,booktabs}
\usepackage[dvipsnames]{xcolor}
\usepackage{rotating,tabularx}

\newcommand{\mc}[1]{\multicolumn{1}{c}{#1}}
\newcommand{\planck}{\textit{Planck}} 
\newcommand{\nHI}{N_{\textsc{HI}}}
\newcommand{\cl}{\mathcal{C}_{\ell}}
\newcommand{\dl}{\mathcal{D}_{\ell}}
\newcommand{\rl}{\mathcal{R}_{\ell}}
\newcommand{\lcdm}{\Lambda\mathrm{CDM}}
\newcommand{\muK}{\mu\mathrm{K}}

\begin{document}

\title{No evidence for dust $B$-mode decorrelation in \planck\ data}
\author{Christopher Sheehy}
\email{csheehy@bnl.gov}
\author{An\v{z}e Slosar}
\affiliation{Physics Department, Brookhaven National Laboratory, Upton, NY 11973}
\date{\today}

\begin{abstract}

  Constraints on inflationary $B$-modes using Cosmic Microwave Background
  polarization data commonly rely on either template cleaning of cross-spectra between maps at different
  frequencies to disentangle galactic foregrounds from the cosmological
  signal. Assumptions about how the foregrounds scale with frequency are
  therefore crucial to interpreting the data. Recent results from the
  \planck\ satellite collaboration claim significant evidence for a
  decorrelation in the polarization signal of the spatial pattern of galactic dust between
  353~GHz and 217~GHz. Such a decorrelation would suppress power in the cross spectrum between
  high frequency maps, where the dust is strong, and lower frequency maps, where
  the sensitivity to cosmological $B$-modes is strongest. Alternatively, it
  would leave residuals in lower frequency maps cleaned with a template derived
  from the higher frequency maps. If not accounted for, both situations would result in
  an underestimate of the dust contribution and thus an upward bias on
  measurements of the tensor-to-scalar ratio, $r$.  In this paper we revisit
  this measurement and find that the no-decorrelation hypothesis cannot be
  excluded with the \planck\ data. There are three main reasons for this: i)
  there is significant noise bias in cross spectra between \planck\ data splits
  that needs to be accounted for; ii) there is strong evidence for unknown
  instrumental systematics whose amplitude we estimate using alternative
  \planck\ data splits; iii) there are significant correlations between
  measurements in different sky patches that need to be taken into account when
  assessing the statistical significance. Between $\ell=55-90$ and over $72\%$
  of the sky, the dust $BB$ correlation between 217~GHz and 353~GHz is
  $1.001^{+.004/.021}_{-.004/.000}$ ($68\%~stat./syst.$) and shows no
  significant trend with sky fraction.
\end{abstract}

\keywords{Interstellar medium: dust -- Submillimeter: ISM -- Polarization --
Cosmic background radiation}

\maketitle
\section{Introduction}

Detection of the primordial $B$-mode signal in the polarization of the Cosmic
Microwave Background (CMB) would imply the existence of tensor modes in the
primordial curvature fluctuations and would be enormously informative in terms
of primordial inflationary
physics~\cite{polnarev85,seljak97b,kamionkowski97,seljak97a}. The
experimental situation is challenging, however, even assuming a perfect instrument: at any
one frequency, the signal of interest is contaminated with foregrounds. The two
main foregrounds are synchrotron radiation at low frequencies and thermal dust
emission at high frequencies. The foregrounds have different spectral indices
compared to the CMB and this allows one to separate them from the signal of
interest. It is often assumed that high frequency maps provide a high signal-to-noise
template of the dust contamination at lower frequencies.

The \planck\ satellite collaboration recently released a paper~\cite[hereafter PIPL]{PIPL}
in which they find evidence for significant amounts of
decorrelation in the $B$-mode signal at $\ell=50-160$ between their 217~GHz and 353~GHz maps.
In other words, the cross-correlation coefficient between $B$-mode polarization in
these two maps

\begin{equation}
  \rl^{BB} =
  \frac{\cl^{BB}(353\times217)}{\sqrt{\cl^{BB}(353\times353)\cl^{BB}(217\times217)}}
\end{equation}

\noindent is less than unity on degree scales. This implies that the two maps
are not simply scaled versions of each other.  In practice, this means that the
map at 353~GHz cannot be used as a template for the dust contribution at lower
frequencies without marginalizing over uncertainty in the assumed degree of
correlation. PIPL also reports a significant trend to more decorrelation at high
galactic latitudes.

This observation is qualitatively consistent with a physical model of how dust
polarization is generated by interaction of dust grains with the galactic
magnetic field~\citep{PIPXLIV,Tassis2015} -- some amount of decorrelation is
expected given variations in the polarization angle and temperature of dust
clouds along the line of sight.  Nevertheless, the amount of decorrelation
reported by \planck\ is surprisingly high. If applied to polarization, the
spatial variations of unpolarized dust temperature ($T_d$) and spectral index
($\beta_d$)~\citep{P2013XI} produce decorrelation that is below the noise floor of the
current data. (In polarization the spatial variations of these parameters are
not measured with statistical significance.)  Using \planck\ data and stellar
extinction measurements, \cite{Poh2016} estimates that decorrelation should
produce a bias on $r$ of $\sim0.0015$ when extrapolating from 353~GHz to
150~GHz. In contrast, PIPL reports that a bias of $r=0.046$ would occur in the
BICEP/\planck\ joint analysis~\citep{BKP} from the level of decorrelation they
measure, a flat $\rl^{BB}=0.95$ between $\ell=50-160$, and possibly much higher
if the trend to higher decorrelation in smaller sky fractions is taken at face
value.  If true, it would have major implications for future $B$-mode surveys
such as CMB Stage IV~\citep{CMBS4}.  In particular, it would drive survey
optimization towards a larger number of more closely spaced frequency
bands. Both of these design choices would likely drive up the cost of these
experiments. This problem therefore warrants further scrutiny. In this paper we
aim to reproduce the results in PIPL and to dig further into data to better
understand the measurement and associated biases.

In the paper we will continually refer to PIPL in order to stress similarities
and differences with their analysis. In particular, we state all our analysis
choices in detail, because these often matter to a surprising degree and to aid
full reproducibility of the results presented in this paper. The paper is
structured as follows. In Section~\ref{sec:data} we discuss the data,
simulations and sky-cut choices used in this work, while in Section
\ref{sec:powerspectra} we show the basic power spectrum results and note the
presence of correlated noise.  In Section~\ref{sec:decorr} we study the
decorrelation coefficient. In Section~\ref{sec:systematics} we examine at what
level systematics known to be present in the \planck\ data could affect the
results. In Section~\ref{sec:skyfrac} we study how the cross-correlation
coefficient varies with the sky fraction, assess the overall statistical
significance of the data, and present the maximum likelihood values for
$\rl^{BB}$.  We conclude in Section~\ref{sec:conclusions}.

\section{Data}
\label{sec:data}

We use the publicly available \planck\ High Frequency Instrument (HFI) data at
217~GHz and 353~GHz~\citep[hereafter Planck 2015 VIII]{P2015VIII}. As in PIPL,
we use two splittings of the data with nominally independent noise to construct
cross spectra that are unbiased by noise. We use the so-called ``detector-set''
splits (hereafter DS) and half-mission splits (hereafter HM).  The HM split
consists of two independent maps constructed from the first and second temporal
halves of the \planck\ nominal mission. The DS split consists of two maps
constructed from the \planck\ full mission data constructed from independent
sets of detector pairs.  Because of our use of the full mission DS split rather than the
nominal mission DS split, the DS split contains additional data compared to the
HM split and therefore has lower noise. This is in contrast to PIPL where the DS
split appears to have the same noise as the HM split, indicating use of of the
nominal mission DS split. Our results using the HM
split are therefore directly comparable to PIPL while our results using the DS
split are not.

In addition to using the HM and DS splits to derive the main results, we also
use additional splits to assess the level of systematics in the data. The
half-ring (HR) split co-adds temporally interleaved hour long time
periods. Systematics that vary over time periods longer than this are thus
common to both halves. We also use HM/DS splits, which are co-added over a single
detector set and a single half-mission. There are therefore four such split
maps, HM$_i$DS$_j$, where $i=[1,2]$.

We use the publicly available PIPL combined galaxy and point source
mask\footnote{\texttt{COM\_Mask\_Dust-diffuse-and-ps-PIP-L\_0512\_R2.00.fits}}
which defines the 9 regions used in the PIPL analysis. This mask defines six
nested regions thresholded on the Planck 857~GHz intensity map that retain
regions of sky defined over $f_{\mathrm{sky}} = 0.2$ to $0.8$ in steps of
$0.1$. After point source masking and apodization the ``large retained'' (LR)
regions are left. They are labeled LR16, LR24, LR33, LR42, LR53, LR63, and LR72,
where the numbers denote the net effective sky coverage as a percentage,
i.e. $100 f_{\mathrm{eff}}^{\mathrm{sky}}$. All of these LR regions overlap each
other. Additionally, the LR63 region is split into its northern and southern
galactic hemisphere halves and labeled LR63N and LR63S. These do not overlap
each other.

\subsection{Simulations}
\label{sec:sims}

\begin{figure*}[!ht]
  \begin{center}
    \begin{tabular}{c}
      \includegraphics[width=1\linewidth]{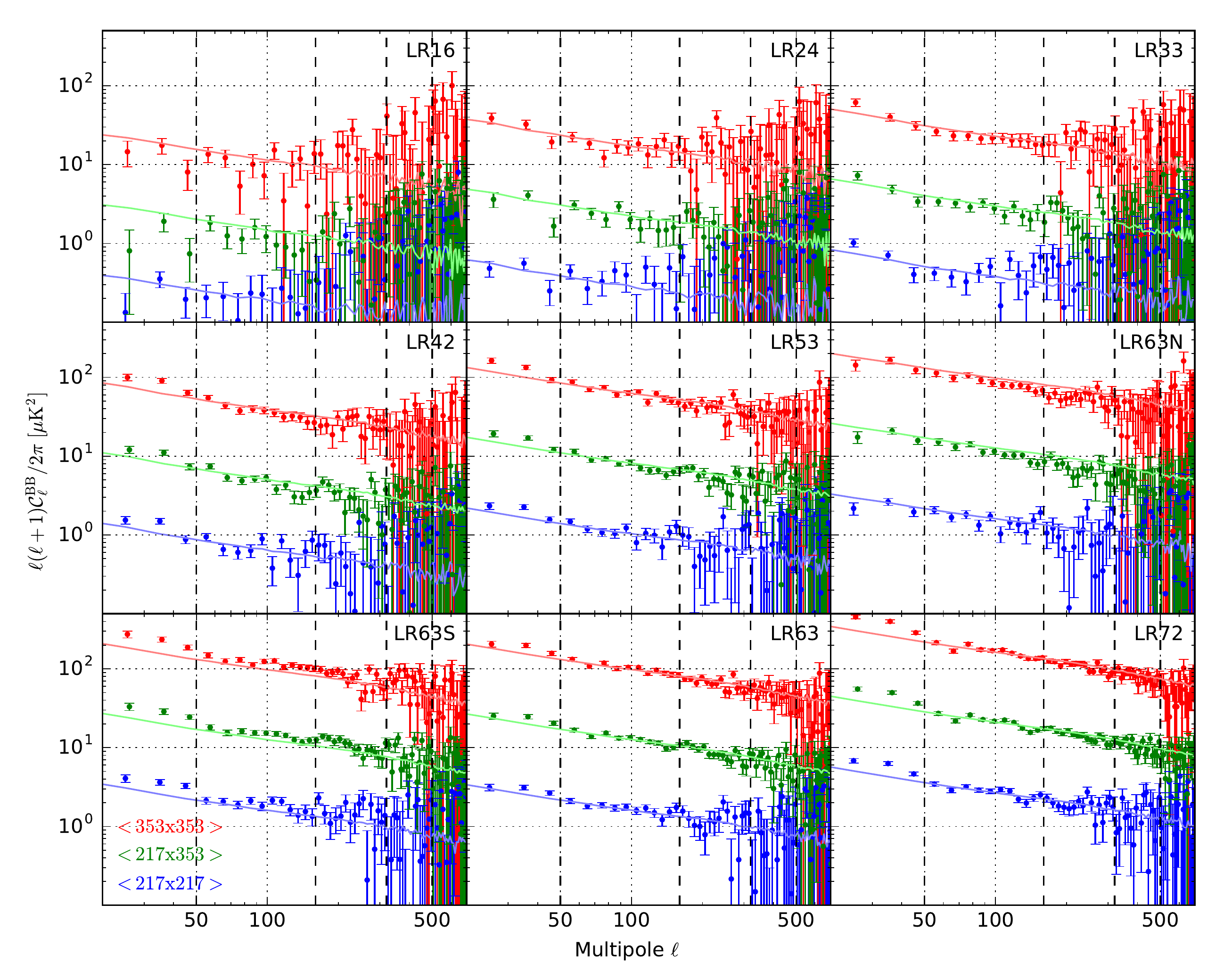} 
    \end{tabular}
  \end{center}
  \caption[example] { \label{fig:rawspec} Half-mission cross spectra on the PIPL
    LR regions in bins of $\Delta\ell=10$. The points are the real data. Error
    bars are the standard deviation of the signal+noise simulations. Solid
    lines are the mean of the simulations. Vertical dashed lines indicate the
    bin edges used in PIPL. Bandpowers for $\ell=20-50$ are plotted for
    completeness but are not used in PIPL.}
\end{figure*}

We construct 500 simulated sky maps for each data split at both 217~GHz and
353~GHz following the procedure outlined in PIPL.  We generate noiseless,
Gaussian realizations of galactic dust plus CMB using the \texttt{synfast}
routine of the \texttt{healpy}\footnote{http://healpix.sf.net} wrapper to
the HEALpix sky pixelization library~\citep{healpix05}.  The input dust power spectra are power
laws with spectral index $\alpha_{EE}=\alpha_{BB}=-2.42$ and amplitudes obeying
the parameters listed in Table 1 of~\citep[hereafter PIPXXX]{PIPXXX}.  Each dust
realization scales in frequency as a modified blackbody with $T_d = 19.6$~K and
$\beta_d=1.59$ \citep{PIPXVII,PIPXXII}. Because the LR16 region was defined
specially for the analysis in PIPL and does not have corresponding parameters in
PIPXXX, we linearly extrapolate the dust amplitude as a function of $f_{sky}$ to
0.2 to find $A^{EE}=25.0~\muK^2$. (Extrapolating as function of neutral hydrogen
column density $\nHI$ yields $A^{EE}=28.0~\muK^2$.) We assume the same $BB/EE$
ratio as LR24, $A^{BB}/A^{EE}=0.49$.

The input CMB power
spectra are generated with the CAMB software\footnote{http://camb.info/} using
the best fit $\lcdm$ model from \cite{P2013XVI}. (Using the more recent 2015
cosmological parameters from \planck\ makes negligible difference.)  The lensing
$B$-mode~\citep{zaldarriaga98} is included by setting the input $BB$ power spectrum to its expected
value and, as such, does not contain off-diagonal power. This is unimportant for
the current study. As in PIPL, we produce independent dust and CMB realizations
for each of the LR regions. The realization is held fixed between
frequencies, with only the dust amplitude changing.

We construct alternative dust simulations using the PySM software, a simple
Python implementation of the \planck\ Sky Model (PSM)~\citep{PSM2013,
  PySM2017}. We run the software with dust models 1 and 2. Dust model 1 scales
the dust template in frequency according to the $\beta_d$ and $T_d$ maps
measured from unpolarized \planck\ data. Dust model 2 scales the dust template
in frequency using a $\beta_d$ map that is drawn from a Gaussian of $\mu=1.59$
and which, after smoothing on degree scales, has $\sigma = 0.2$. Both dust model
1 and dust model 2 are reported in~\cite{PySM2017} to be consistent with
\planck\ data. Dust model 2 produces much more decorrelation than dust
model 1 (see Figure~\ref{fig:Rfine_LR63}). We note that there is only a single
PySM dust realization, which remains fixed between realizations and LR regions
in these simulations.

Also as in PIPL, we construct noise realizations as random Gaussian realizations
of $Q$ and $U$ maps obeying the 4x4 $QU$ covariance matrix which
\planck\ provides for every map pixel of every data split. By construction in
these simulations, noise between pixels and between data splits is
uncorrelated. We produce 500 full sky noise realizations and hold these fixed
between LR regions so that, as in the data, there is a common component to the
noise in each of the nested LR regions. This differs from PIPL, which reports
that they produce independent noise realizations for each of the LR regions.

Additionally, we use the Full Focal Plane Monte Carlo noise simulations, namely
the FFP8 simulations \citep{P2015XII}, as an alternative to the $QU$ covariance noise
simulations for the HM split. PIPL reports that they did not use the FFP8 maps
directly as input realizations because of the presence of instrumental noise in
the polarized dust component of the PSM.  The FFP8 Monte Carlo noise
realizations contain only simulated noise, however, and are unaffected by this
issue. PIPL also reports that the FFP8 noise realizations that are available
through the \planck\ Legacy
Archive\footnote{\texttt{http://www.cosmos.esa.int/web/planck/pla}} (PLA) are in
good agreement with the noise realizations constructed from the corresponding
$QU$ covariance maps. Though we confirm this agreement, only the full mission,
full detector set FFP8 noise simulations are available through the PLA. Because
both this analysis and that in PIPL use cross spectra between data splits, the
publicly available FFP8 noise simulations are not directly useful for the
current analysis. We therefore requested and obtained 1000 realizations of the HM split FFP8 noise
simulations at 217~GHz and 353~GHz and 100 realizations at other frequencies. These have subsequently been made publicly
available via the \planck\ data archive hosted on
NERSC.\footnote{http://crd.lbl.gov/departments/computational-science/c3/} 

Unlike the $QU$ covariance noise simulations, the FFP8 simulations are expected
to reproduce any noise correlations between data splits produced by the
\planck\ map making procedure. We therefore use
the FFP8 noise realizations in Section~\ref{sec:noisecorr} to determine any
noise bias present in the data splits. All other results presented in this paper
use the $QU$ covariance noise realizations.

\subsection{Map preparation}

The publicly available PIPL mask is provided at a HEALpix resolution of
Nside=512. The publicly available Planck HFI maps are provided at the higher
resolution of Nside=2048. PIPL does not describe their procedure for either
downgrading the resolution of the HFI maps or upgrading the resolution of the
mask. We opt to downgrade the HFI map resolution by averaging the sets of
Nside=2048 pixels that form an Nside=512 pixel. (We use the \texttt{healpy}
\texttt{ud\_grade} utility.) The full resolution FFP8 noise realizations contain
a number of \texttt{unseen} entries in the HM split that are not present
in the data maps. We therefore mask these pixels and exclude them from the
average in every map at every frequency prior to downgrading. This has the
effect of increasing the effective noise in those Nside=512 pixels that
contained an \texttt{unseen} in the superset of Nside=2048 pixels that constitute it. We
fully account for this by generating the noise realizations at the native
resolution of Nside=2048 and applying the mask in the same manner as we
apply it to the real data.

We generate realizations of noiseless $a_{lm}$'s for CMB and dust and add them
together with the appropriate frequency scaling for 217~GHz and 353~GHz (see
Section~\ref{sec:sims}). We then multiply the $a_{lm}$'s by the Gaussian beam window
function appropriate for the HFI 217 and 353 GHz beams (4.99 and 4.82 arcmin
FWHM, respectively). We also multiply the $a_{lm}$'s by the HEALpix pixel window
function appropriate for an Nside=512 map to capture power suppression from
binning into pixels. Any effects from not generating the maps at
the full resolution and applying the \texttt{unseen} mask are restricted to the pixel scale ($\ell\sim1500$) and are
thus irrelevant for the subsequent analysis. We then add the signal realizations
and downgraded noise realizations to produce the final simulated maps
(referred to as signal + noise simulations).

\subsection{Power spectrum estimation}

As in PIPL, we use the XPol power spectrum estimator~\citep{Xpol2005} to derive
our main results. We also obtain similar results with the
PolSpice\footnote{http://www2.iap.fr/users/hivon/software/PolSpice/}
estimator~\citep{Pspice2004}. Both estimators correct for $EB$ mixing in the
mean resulting from incomplete sky coverage. PolSpice does no binning in $\ell$
and returns $\cl = \left<a_{\ell m}a_{\ell m}^*\right>$. XPol requires the user
to specify multipole bins and returns $\dl = \ell(\ell+1)\cl/2\pi$. In broad
bins, we obtain consistent results between the two estimators only when we
multiply PolSpice $\cl$s by $\ell(\ell+1)/2\pi$ (to make them $\dl$s) prior to
binning. In the main results derived in this work using the HM and DS splits, we
specify bins of width $\Delta\ell=5$ to XPol and re-bin these spectra to produce
broader bins. We find this gives consistent results compared to specifying the
broad bins to XPol directly. Binned $\dl$'s are referred to as ``bandpowers.''
In Section~\ref{sec:systematics}, we compute the unbinned spectra of additional
splits to assess the level of systematics in the data. For these results we use
the PolSpice estimator.

We apply the estimator to the signal + noise simulations and to the real
data. We also apply it to the signal and noise simulations separately. At each
frequency, we compute the cross spectrum between the HM or DS split halves:

\begin{equation}\label{eq:intrafreq}
\cl^{XX} (\nu \times \nu) = 
\cl^{XX} (\mathrm{map}_{\nu}^1 \times \mathrm{map}_{\nu}^2 )
\end{equation}

\noindent where $XX \in {EE, BB}$ and the superscripts $1$ and $2$ denote the
split half. As in PIPL, we also compute the cross spectrum between frequencies by
taking the mean of all four independent crosses:

\begin{equation}\label{eq:interfreq}
\cl^{XX} (\nu_1 \times \nu_2) = \frac{1}{4} \sum_{i,j}
\cl^{XX} (\mathrm{map}_{\nu_1}^i \times \mathrm{map}_{\nu_2}^j )
\end{equation}

\noindent where $i$ and $j$ take the values 1 and 2, representing the two
independent splits. As long as the noise is uncorrelated between all four
$\mathrm{map}_{\nu}^i$, the cross spectra computed with Eqs.~\ref{eq:intrafreq}
and \ref{eq:interfreq} have no additive noise bias.

\section{Power spectrum Results}
\label{sec:powerspectra}

Figure~\ref{fig:rawspec} shows the $BB$ intra- and inter-frequency cross spectra
on each LR region computed from the HM split using Eqs.~\ref{eq:intrafreq} and~\ref{eq:interfreq}. The spectra are binned into
bandpowers of width $\Delta\ell= 10$. The agreement between the real data and
the mean of simulations indicates the appropriateness of our simulations for
modeling the real data. (Ultimately, this indicates the appropriateness of the
dust power law parameters in Table 1 of PIPXXX for describing the galactic dust
in these regions of sky.) We attribute the slight deviation of the real data
from the simulations in the LR63S and LR63N regions to the fact that dust power
law parameters are only available for the full LR63 region and would be somewhat
different if fit for separately on the north and south patches. 

The DS split power spectra (not shown) are consistent over all multipoles to
within twice the total error bars shown in Figure~\ref{fig:rawspec}, indicating
the relative unimportance of instrumental systematics for measuring quantities
affected by dust sample variance, such as the dust power law
parameters.
We do not have the DS split FFP8 noise
simulations and therefore cannot assess the consistency of the HM and DS
bandpowers to the level of instrumental noise. 
We therefore cannot test for HM-DS data consistency at the level
required for super-sample variance measurements, such as the decorrelation parameter
introduced in the next section.

\subsection{Intra-frequency noise correlations}
\label{sec:noisecorr}

\begin{figure}[]
  \begin{center}
    \begin{tabular}{c}
      \includegraphics[width=1\columnwidth]{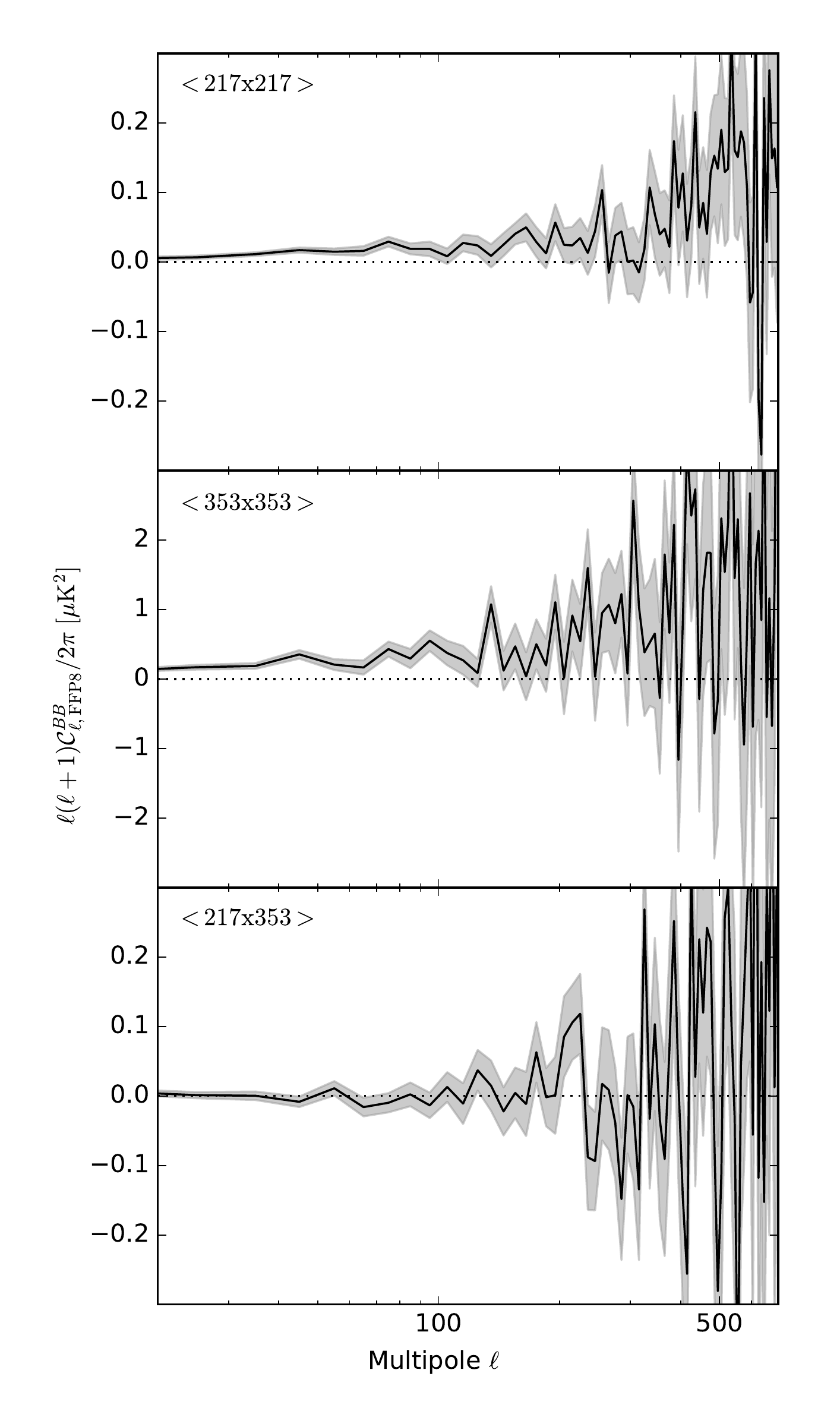} 
    \end{tabular}
  \end{center}
  \caption[example] { \label{fig:noispec} Half-mission cross spectra of the FFP8
    Monte Carlo noise simulations in the LR63 region. The binning is the same as
    in Figure~\ref{fig:rawspec}. The solid line is the mean over FFP8
    realizations and the shaded region is the standard error, computed as the
    standard deviation over realizations divided by the square root of the the
    number of realizations.}
\end{figure}

With the FFP8 HM noise simulations we can test for noise bias in the power
spectra shown in Figure~\ref{fig:rawspec}.
Figure~\ref{fig:noispec} shows the HM cross spectra of the FFP8 HM noise
simulations on the LR63 region. These spectra indicate significant positive bias
in $217\times217$ and $353\times353$, and no bias in $217\times353$. The bias ranges from
$\sim1\%$ of the dust signal at $\ell=50$ to $\sim15\%$ at $\ell=700$ in
$217\times217$. The bias is similar in other LR regions but is measured with somewhat
less statistical precision.  There is no measured bias in the $QU$ covariance
generated noise simulations. 

Such a bias in the \planck\ data is expected given the destriping procedure
described in Planck 2015 VIII, Section 6.5, which describes the trade-off in
accuracy vs. noise correlation given the choice of using a baseline offset
computed for each subset independently (lower accuracy but maintains independent
noise), or using the full frequency, full mission baselines to destripe the
subset halves (higher accuracy at the cost of introducing noise correlations).
The Planck 2015 HFI data release uses the latter destriping procedure. The text
of Planck 2015 VIII explicitly states that full mission destriping introduces
noise correlations between detector set maps, and Figure~17 of that paper shows the FFP8
detector-set $EE$ noise cross spectrum at 100~GHz, which
peaks at $\cl \sim 0.0035~\muK^2$ at $\ell=2$ and falls steeply with $\ell$,
though still appears visibly positive at $\ell=50$. (In Figure~\ref{fig:noispec}
of this work, the bias appears to increase with $\ell$ because of the $\ell^2$ scaling.)
This amplitude matches the noise correlation we observe in the 100~GHz
half-mission FFP8 noise cross spectrum (not shown). We have verified that the destriping
procedure produces similar correlations in the half-mission split noise cross
spectra, and that the FFP8 noise simulations include these induced noise correlations (private communication, J. Borrill).  We therefore conclude that the
noise correlation shown in Figure~\ref{fig:noispec} is present in both the
Planck 2015 HM and DS split maps. We note that PIPL does not account for any
bias introduced by noise correlations.

Assuming that the signal is the same in each data subset and that it is
uncorrelated with noise, the total measured cross spectrum is
$ \left<(S+N_1)(S+N_2)\right> =\left<S^2\right> + \left<N_1
  N_2\right>$
where $S$ is the signal and $N_i$ is the noise in each subset.
Figure~\ref{fig:noispec} shows the relatively small but important
correlated noise term $\left<N_1 N_2\right>$, which must be subtracted
(i.e. ``debiased'') from the measurement.

\section{Decorrelation}
\label{sec:decorr}

\begin{figure}[]
  \begin{center}
    \begin{tabular}{c}
      \includegraphics[width=1\columnwidth]{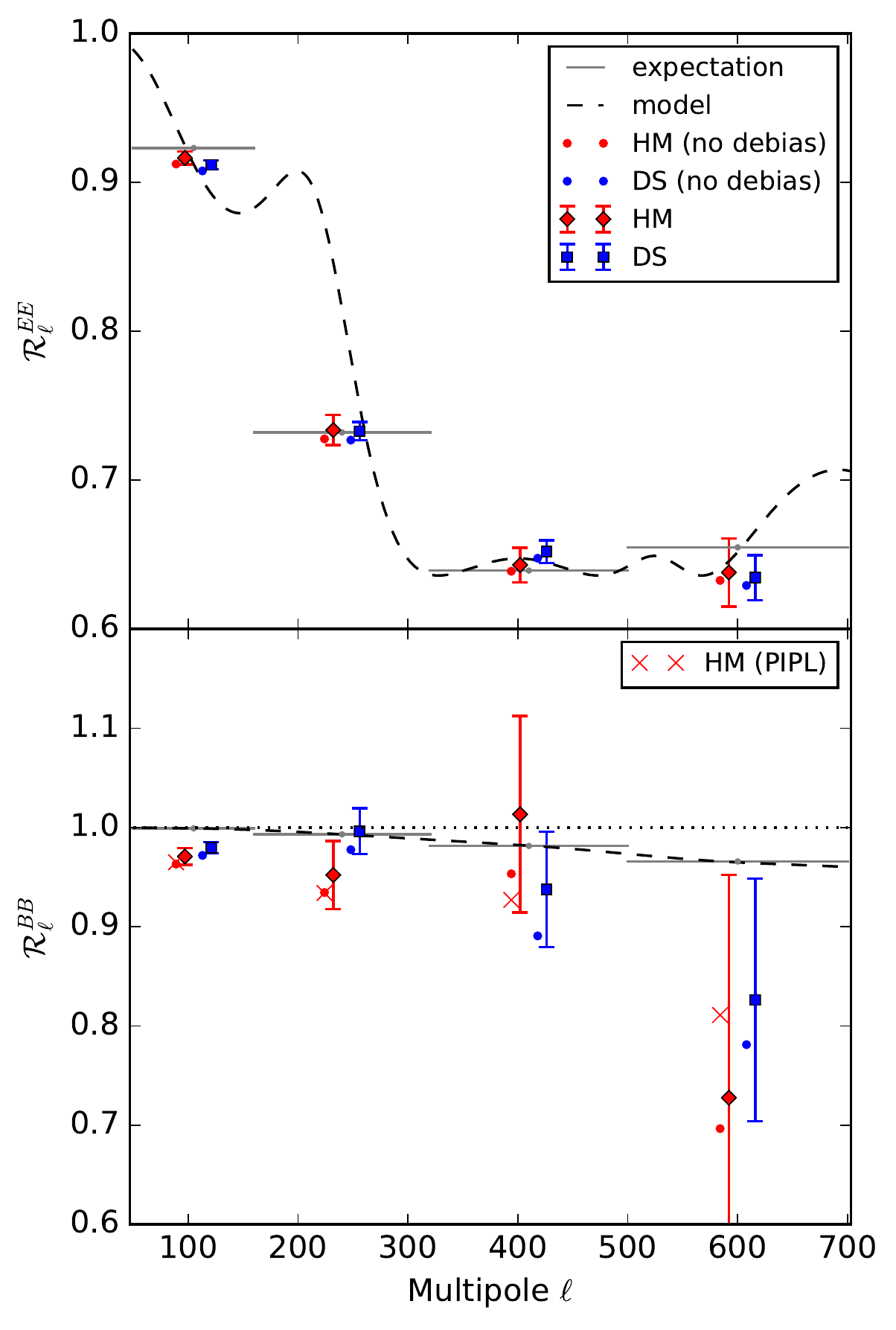} 
    \end{tabular}
  \end{center}
  \caption[example] { \label{fig:LR63spec} Correlation ratios $\rl^{EE}$ (top
    panel) and $\rl^{BB}$ (bottom panel) in the LR63
    region. Correlation ratios calculated from the HM and DS splits are shown as
    the red diamonds and blue squares, respectively. The red and blue dots are
    the correlation ratios computed from the same spectra but without debiasing
    the expected noise correlation.  The red x's in the bottom panel are
    $\rl^{BB}$ reported in the appendix of PIPL, which are directly comparable to the red dots.
    The dashed black line is the expected correlation ratio given the relative
    amplitudes of dust and CMB in the LR63 region. The model expectation values
    are shown as gray horizontal lines indicating the bin. The error
    bars are computed as the median absolute deviation of the signal+noise
    simulations.}
\end{figure}

\subsection{Correlation ratio}

We compute the correlation ratio between 217 and 353~GHz, defined in PIPL as

\begin{equation}
\label{eq:Rdef}
  \rl^{XX} = 
  \frac{\cl^{XX}(353\times217)}{\sqrt{\cl^{XX}(353\times353)\cl^{XX}(217\times217)}}
\end{equation}

\noindent where $XX \in {EE,BB}$. Eq.~\ref{eq:intrafreq} is used to compute the
two terms in the denominator and Eq.~\ref{eq:interfreq} is
used to compute the numerator. Any operation which multiplies the $a_{lm}$'s of
a given map by an arbitrary function of $\ell$ cancels in the correlation ratio.
Therefore $\rl$ is unaffected by convolution with a circularly symmetric beam,
multiplication by the pixel window function, or by many calibration errors. (In
principle, the beam window functions for the detector-set cross do not
perfectly cancel in the ratio. We have verified using the HFI beam
window functions provded in the \planck\ Reduced Instrument
Model\footnote{HFI\_RIMO\_R2.00.fits} that the non-cancellation produces
deviations of $\rl<10^{-6}$ at $\ell<700$.)

If there is no noise bias or instrumental systematics and the sky at 217~GHz is
perfectly spatially correlated with the sky at 353~GHz, then $\rl^{XX} = 1$.
Such would be the case if the maps contained a single component with a spatially
invariant spectral energy distributions (SED). If two or more components with
different SEDs contribute to the maps then they deviate from perfect spatial
correlation and $\rl<1$. We expect this decorrelation from the admixture of dust
and CMB. In $BB$, only the lensing $BB$ produces this decorrelation. Since the
lensing $BB$ is small compared to the dust at low $\ell$, the amount of
decorrelation it produces is quite small and relatively immune to assumptions
about the relative power in the two components. 

Lastly, additional decorrelation will be produced if any component contains a
spatially varying SED, for instance, from a spatially dependent $\beta_d$ or
from polarization angle rotations. As noted in the introduction, such effects
are predicted to exist at a small level~\citep{Tassis2015, Poh2016}.

Figure~\ref{fig:LR63spec} is analagous to Figure 2 of PIPL and shows $\rl^{EE}$ and $\rl^{BB}$ for the LR63 region
using the same four bins as PIPL ($\ell=50-160$, $\ell=160-320$, $\ell=320-500$,
and $\ell=500-700$).  As in PIPL, the error bars are computed as the median of
the absolute deviation of the signal + noise simulations. Prior to noise
debiasing, we find nearly exact agreement with PIPL in $\rl^{BB}$ in the first
two bins for the HM split. In the last two bins there are small, $<1\sigma$
shifts in $\rl^{BB}$.  (As stated in Section~\ref{sec:data}, the DS split is not
exactly comparable with PIPL.)  Noise debiasing results in a $\sim 1\sigma$
shift upwards in $\rl^{BB}$ in the first bin for both the the HM and DS splits.

In $\rl^{EE}$ there are two significant differences with the figure in
PIPL. First, we find significantly smaller error bars for $\rl^{EE}$ compared to
$\rl^{BB}$.  This is perhaps because PIPL appears to transfer the $\rl^{BB}$
error bars onto the $\rl^{EE}$ bandpowers.  Second, our model expectation values
differ somewhat from PIPL due to more careful binning. We can reproduce the PIPL
results by binning the $\rl$ model curve computed from unbinned model
spectra. Where $\rl$ is changing rapidly, this can produce significant shifts in
the expectation values compared to binning the constituent spectra, which is the
procedure that is consistent with how the data are treated.  

Because the $\lcdm$ $E$-mode power is similar in amplitude to dust power, we
find that the expected $EE$ decorrelation is significantly affected by the assumed
amplitude of dust power in each $\ell$ bin.  Noiseless simulations run with PySM
modified to produce zero decorrelation show significant deviations from the $EE$
model curve in Figure~\ref{fig:LR63spec}.
Because dust is the only significant
contributor to the $B$-mode power, however, there is almost no dependence of
$\rl^{BB}$ on $\lcdm$ sample variance or the assumed dust
amplitude. Accordingly, the PySM simulations show excellent agreement with the
$BB$ model curve in Figure~\ref{fig:LR63spec}.
Therefore, as in PIPL, we only use $\rl^{BB}$ to derive results.

\subsection{Alternative binning}

\begin{figure}[]
  \begin{center}
    \begin{tabular}{c}
      \includegraphics[width=1\columnwidth]{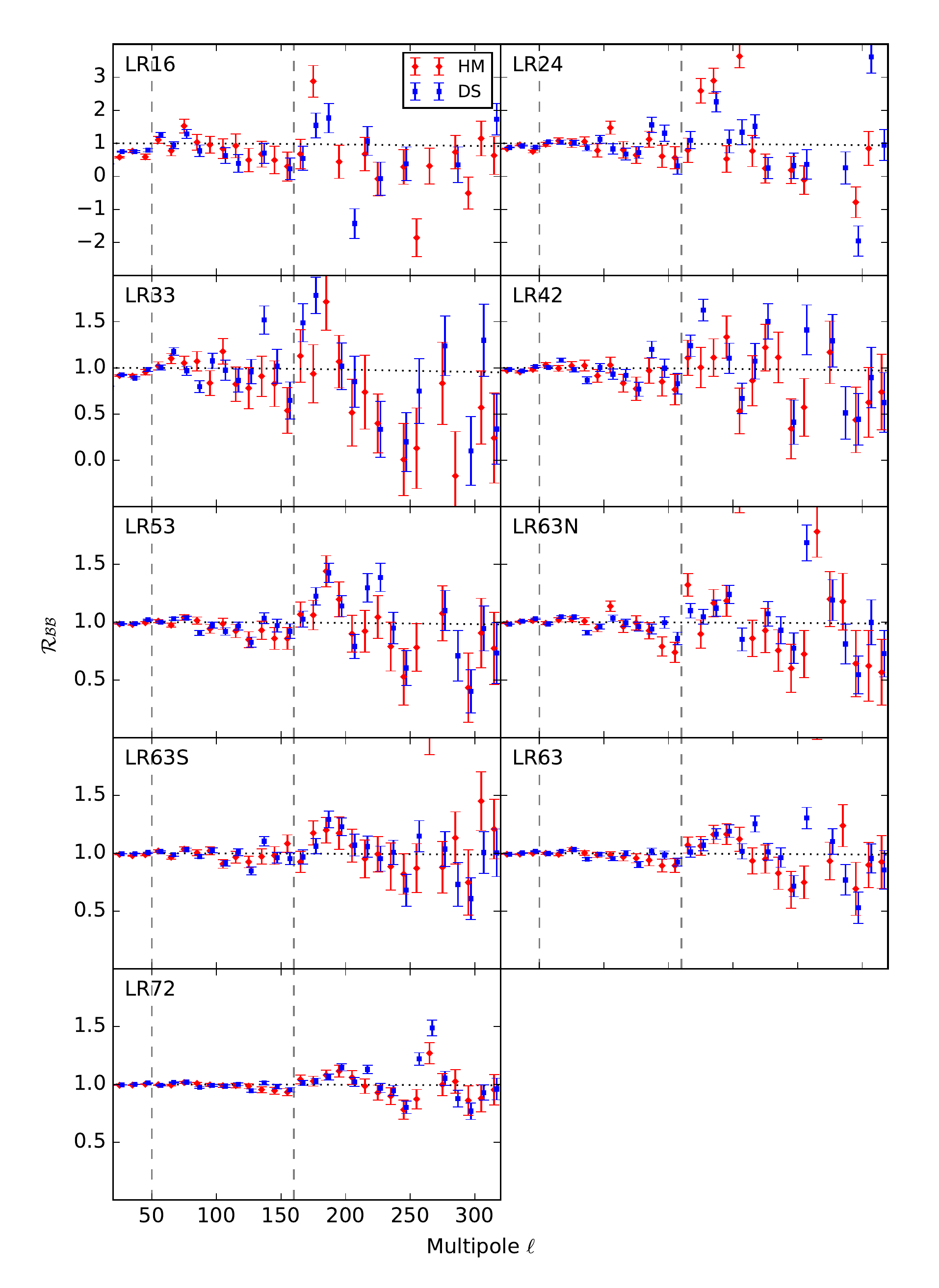} 
    \end{tabular}
  \end{center}
  \caption[example] { \label{fig:Rfine_allLR} Noise debiased $\rl^{BB}$ on all LR
    regions in bins of width $\Delta\ell=10$. The HM and DS splits are shown as
    red and blue points, respectively. The error bars are computed as the
    standard deviation of the signal+noise simulations. The dashed vertical gray
    lines indicate the PIPL bin edges. The dotted line is the model expectation
    value.}
\end{figure}

\begin{figure}[]
  \begin{center}
    \begin{tabular}{c}
      \includegraphics[width=1\columnwidth]{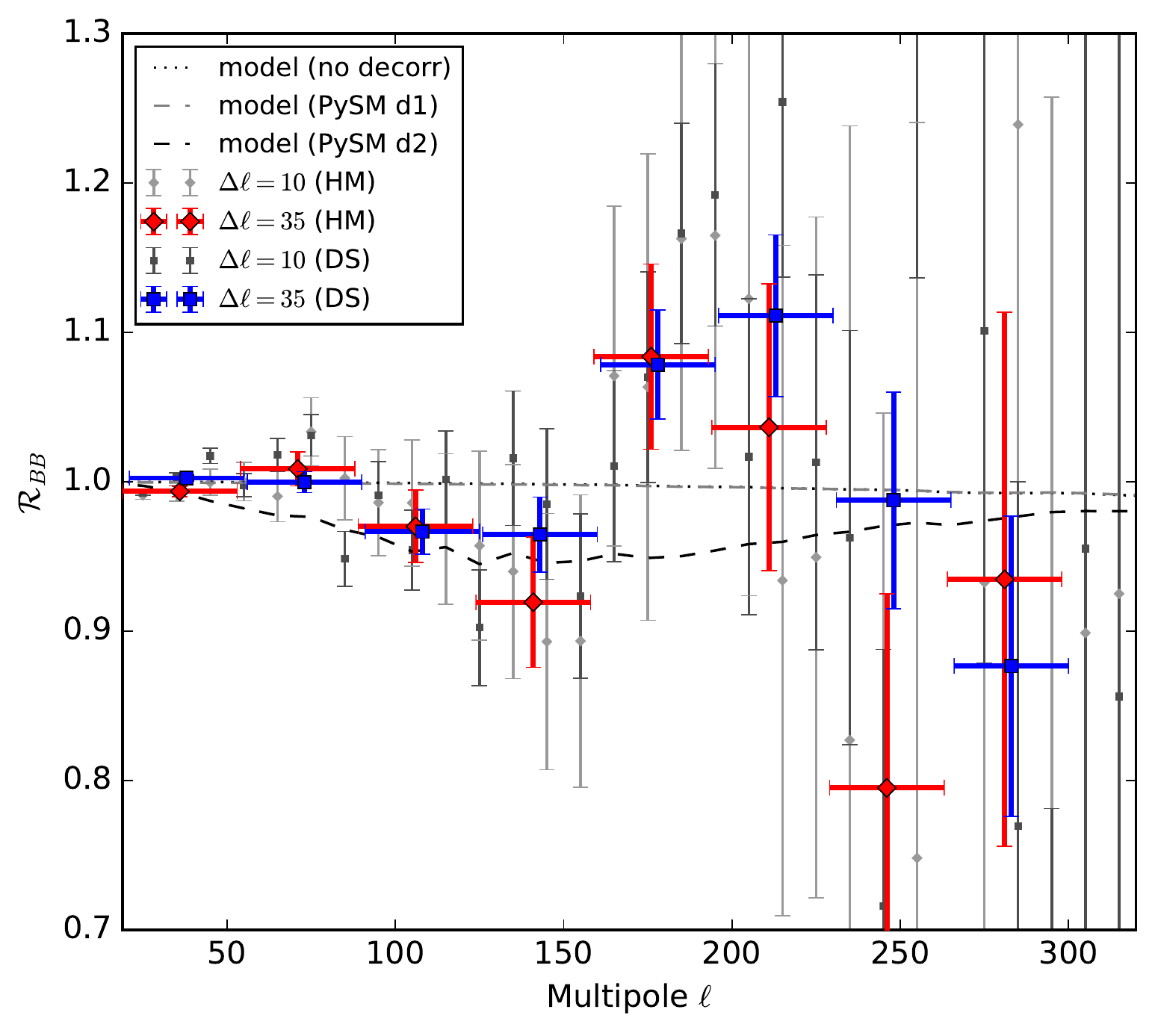} 
    \end{tabular}
  \end{center}
  \caption[example] { \label{fig:Rfine_LR63} Noise debiased $\rl^{BB}$ on the
    LR63 region. The gray points are the same HM and DS data points shown
    in Figure~\ref{fig:Rfine_allLR}. The red diamonds and blue squares use an
    alternative binning of $\Delta\ell=35$ beginning from $\ell=20$. Error bars
    are the standard deviation of the signal+noise sims. The dotted black line
    shows the $\rl^{BB}$ model expectation for no dust decorrelation. The
    dashed gray/black lines show the decorrelation produced by PySM dust model
    1/2. PySM dust model 1 is indistinguishable from the no decorrelation model.} 
\end{figure}

We now compute $\rl^{BB}$ using finer bins than presented in
PIPL. Figure~\ref{fig:Rfine_allLR} shows $\rl^{BB}$ computed from binning $\dl$
in bins of $\Delta\ell = 10$ (i.e. the same spectra shown in
Figure~\ref{fig:rawspec}).  Figure~\ref{fig:Rfine_LR63} shows the LR63 panel,
adding bins of $\Delta\ell = 35$ starting from $\ell=20$. We choose these latter
bins to be the same as the bins used in the BICEP-\planck\ joint analysis. The
$\ell = 55-90$ bin has comparable signal-to-noise as the $\ell=50-160$ bin and
shows no evidence for decorrelation in either the HM or DS splits. It appears
inconsistent with the flat decorrelation assumed by PIPL
($\mathcal{R}_{50-160}^{BB}=0.95$). It also appears inconsistent with PySM dust model
2. It is consistent with both the no-decorrelation model and PySM dust model 1,
which shows negligible additional decorrelation from the no-decorrelation model.

\section{Systematics}
\label{sec:systematics}

\begin{figure}[]
  \begin{center}
    \begin{tabular}{c}
      \includegraphics[width=1\columnwidth]{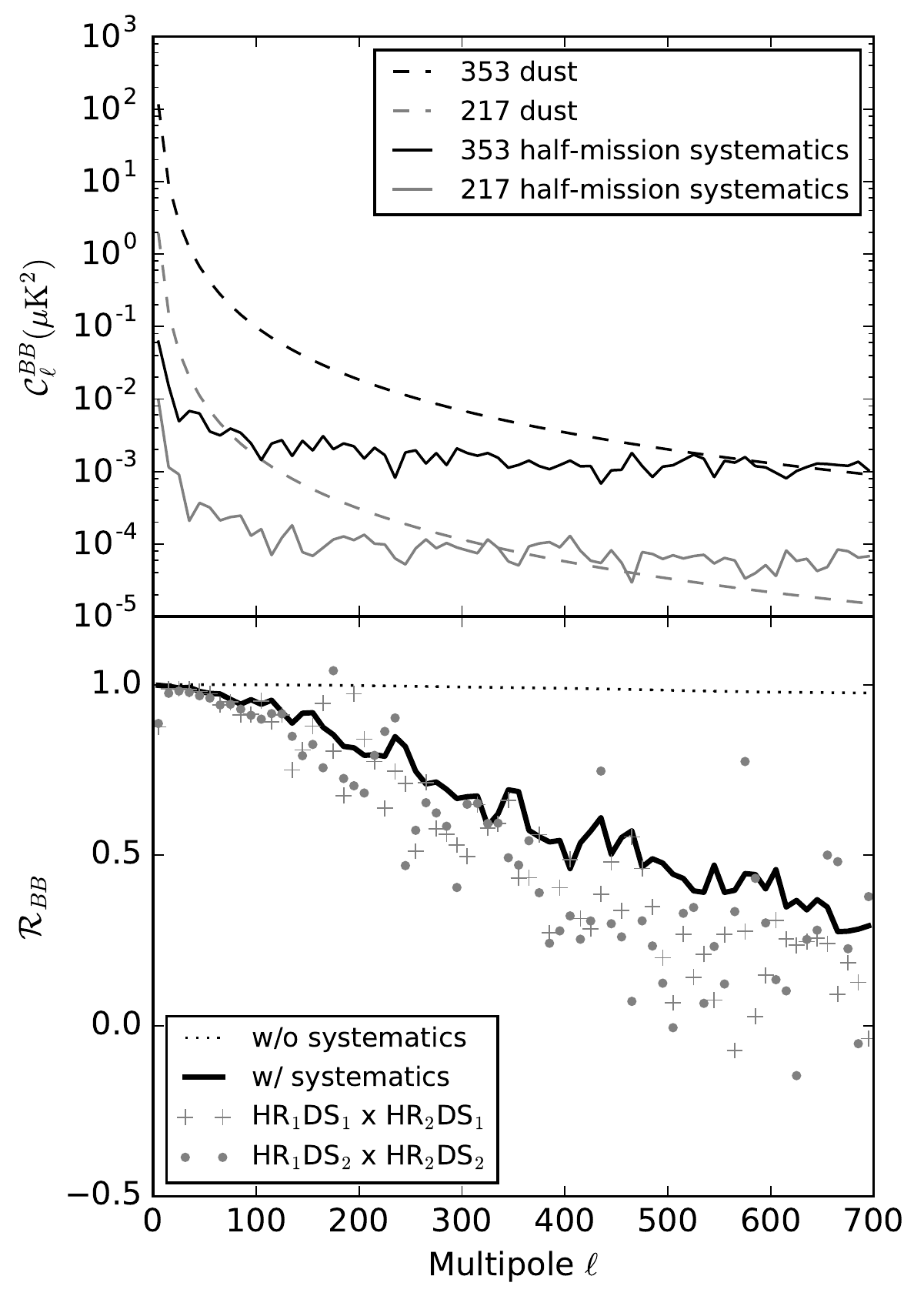}
    \end{tabular}
  \end{center}
  \caption[example] { \label{fig:syst3} \textit{Top panel}: Model dust
    $\cl^{BB}$ for 353~GHz and 217~GHz (dashed lines)
    in LR72, and excess power in the HM map difference null test
    relative to the FFP8 noise simulations (solid lines). \textit{Bottom panel}:
    Predicted $\rl^{BB}$ in LR72 with no decorrelation and no systematics
    (dotted black line) and in the presence of systematics given by the solid
    lines in the top panel (solid black line). The data points are the measured
    $\rl^{BB}$ computed from the half-ring split for each detector set
    separately.}
\end{figure}

\begin{figure}[]
  \begin{center}
    \begin{tabular}{c}
      \includegraphics[width=1\columnwidth]{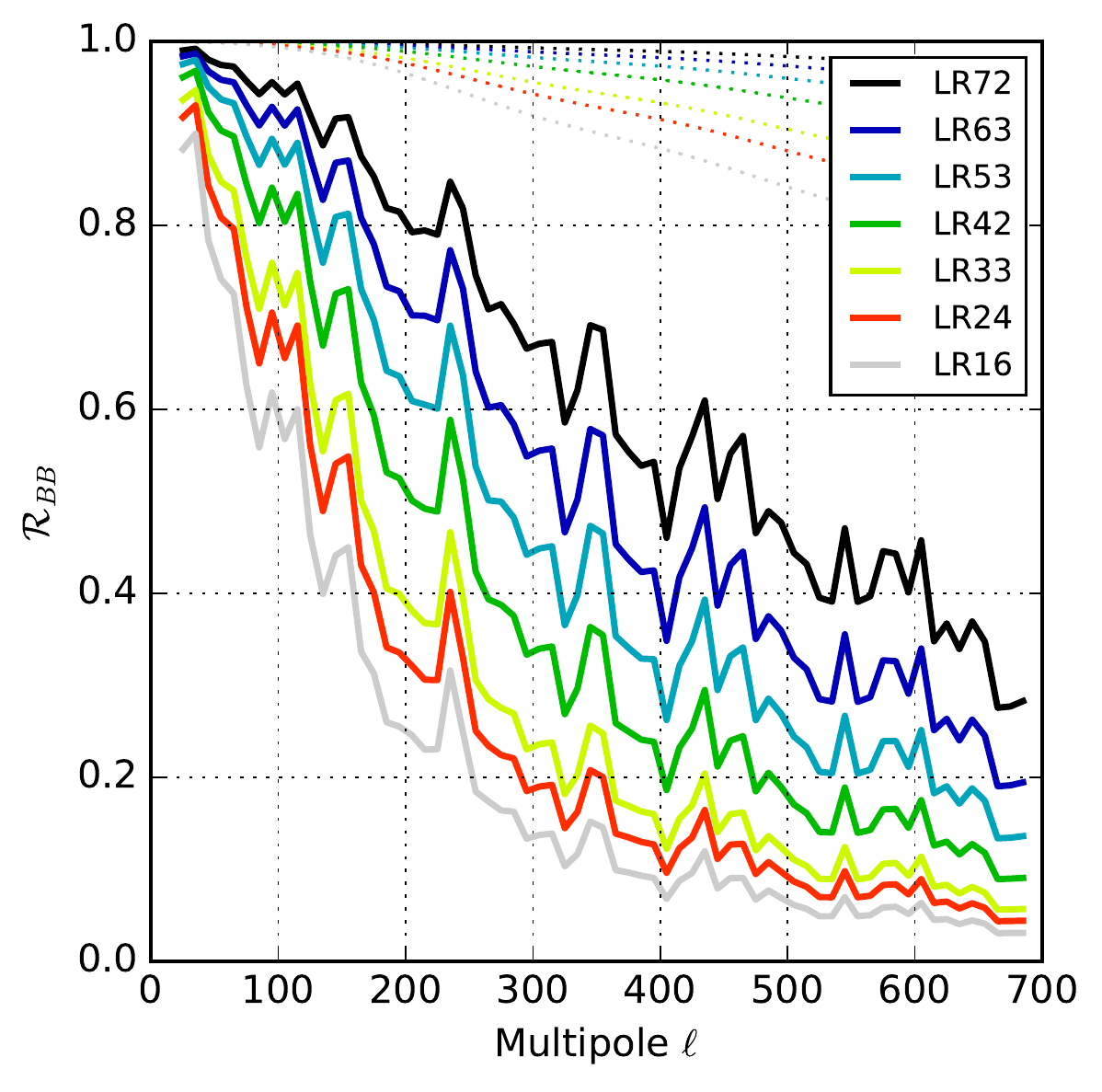}
    \end{tabular}
  \end{center}
  \caption[example] { \label{fig:syst4} Expectation for $\rl^{BB}$ in the
    presence of a correlated systematic of the same amplitude as the uncorrelated
    systematic shown in Figure~\ref{fig:syst3}. The expectation without
    systematics is given by the corresponding dotted lines.}
  
\end{figure}

\begin{figure}[]
  \begin{center}
    \begin{tabular}{c}
      \includegraphics[width=1\columnwidth]{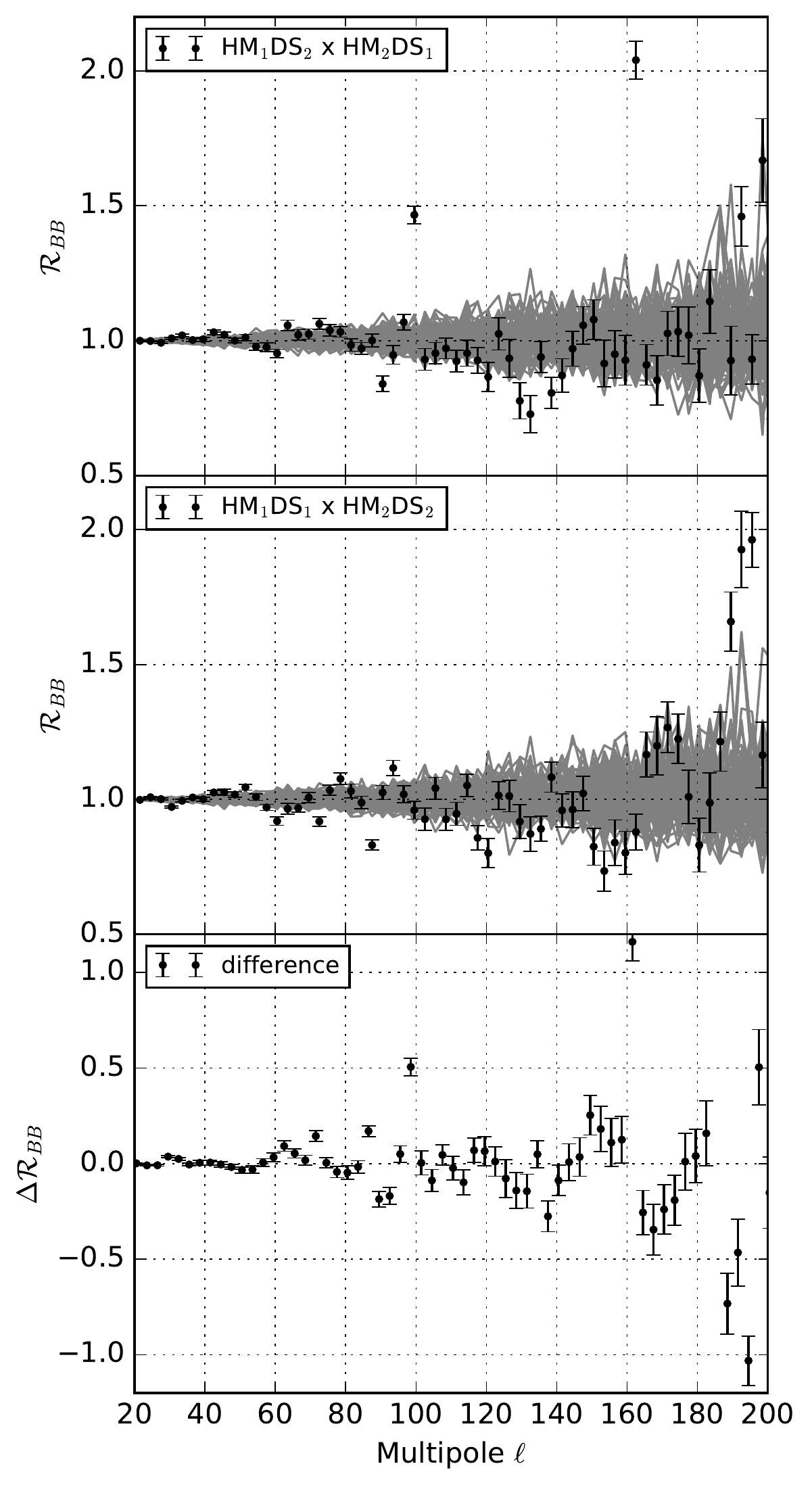}
    \end{tabular}
  \end{center}
  \caption[example] { \label{fig:syst1} \textit{Top and middle panels}:
    $\rl^{BB}$ on LR72 measured in bins of $\Delta\ell=3$ using two independent
    cross spectra of the four HM$_i$DS$_j$ data splits.  The black points are the
    data. The error bars are computed as the standard deviation of 100
    corresponding signal+noise simulations with the signal realization held
    fixed, which are shown as gray lines. Points with error bars should be
    compared to $1$ rather than to the simulations, which are shown only to give
    a visual indication of the distribution. \textit{Bottom panel}: the
    difference, $\Delta\rl^{BB}$, between the data points in the top and middle
    panels, which should be consistent with zero. The error bars are the
    standard deviation of the corresponding differences between simulation
    realizations. }
\end{figure}

The use of cross spectra to calculate the denominator of Eq.~\ref{eq:Rdef} means
that $\rl$ will be biased by the presence of systematics that correlate between
either halves of the HM or DS split or between 217 and 353. Figure 11 of Planck
2015 VIII shows a clear failure of map difference null tests constructed from
single-frequency data splits, $\mathrm{map}^{\mathrm{HM}_1}_{\nu} -
\mathrm{map}^{\mathrm{HM}_2}_{\nu}$ and $\mathrm{map}^{\mathrm{DS}_1}_{\nu} -
\mathrm{map}^{\mathrm{DS}_2}_{\nu}$.  There is significant excess power in the
difference maps compared to power in difference maps of the corresponding FFP8
noise simulations. Planck 2015 VIII attributes this to instrumental systematics,
and a subsequent paper~\citep{P2015XLVI} finds it to be largely the result of
non-linearity in the analog-to-digital converter.

Instrumental systematics that contaminate a map difference null test are by
definition uncorrelated between data halves. As such, they do not bias $\rl$. We
can predict what bias such a systematic would produce were it instead correlated
between data splits.  The top panel of Figure~\ref{fig:syst3} shows the LR72
model dust spectra at 217 and 353 compared to the uncorrelated systematics in
the HM maps, which we compute as the excess power in the HM difference maps
compared to the FFP8 noise simulations:

\begin{eqnarray} 
\cl^{syst} & = & \cl [ ( \mathrm{map}_{\nu}^{\mathrm{HM_1^{data}}} -
                         \mathrm{map}_{\nu}^{\mathrm{HM_2^{data}}})/2 ] - \\
           &   & \left<\cl [ ( \mathrm{map}_{\nu}^{\mathrm{HM_1^{FFP8}}} -
                         \mathrm{map}_{\nu}^{\mathrm{HM_2^{FFP8}}})/2 ]\right> .\nonumber
\end{eqnarray}

\noindent where the expectation value is taken over realizations.

The 217 systematics curve in Figure~\ref{fig:syst3} is
comparable to the difference between the ``Half Mission'' and ``FFP8'' lines in
the bottom panel of Fig. 11 of Planck 2015 VIII. (The main difference is that
the present work shows $BB$ systematics while the \planck\ figure shows $EE$
systematics.) Both figures show excess power in 217 of $\sim 
10^{-2}~\muK^2$ at $\ell=10$ and $\sim 10^{-4}~\muK^2$ at
$\ell=100$. We therefore conclude that the uncorrelated systematics in the 217
HM split maps dominate the LR72 dust signal at $\ell>300$ and are $10\%$ of the
dust signal at $\ell=50$. In the 353 maps, the systematics are fractionally
lower relative to the dust signal but are still $10\%$ at $\ell=150$.

The black line in the bottom panel of Figure~\ref{fig:syst3} shows the
expected bias on $\rl$ that such a systematic would produce in LR72 if it were
instead correlated between data split halves. One such data split that preserves
correlations of instrumental systematics is the HR split.  The data points in
the bottom panel of Figure~\ref{fig:syst3} show $\rl^{BB}$ computed from HR
cross spectra of maps built from individual detector sets. There is a large
downward bias on $\rl^{BB}$ whose magnitude is comparable to the level predicted
from the HM map difference null test. The fact that a bias of this magnitude is
not observed in $\rl$ computed from HM or DS cross spectra indicates that the
portion of instrumental systematics that is correlated between the data split
halves is small compared to the uncorrelated portion. However, if even a small
fraction of this systematic were correlated, it would produce a bias on
$\rl^{BB}$ that is significant compared to the measurement uncertainty.

\begin{figure}[]
  \begin{center}
    \begin{tabular}{c}
      \includegraphics[width=1\columnwidth]{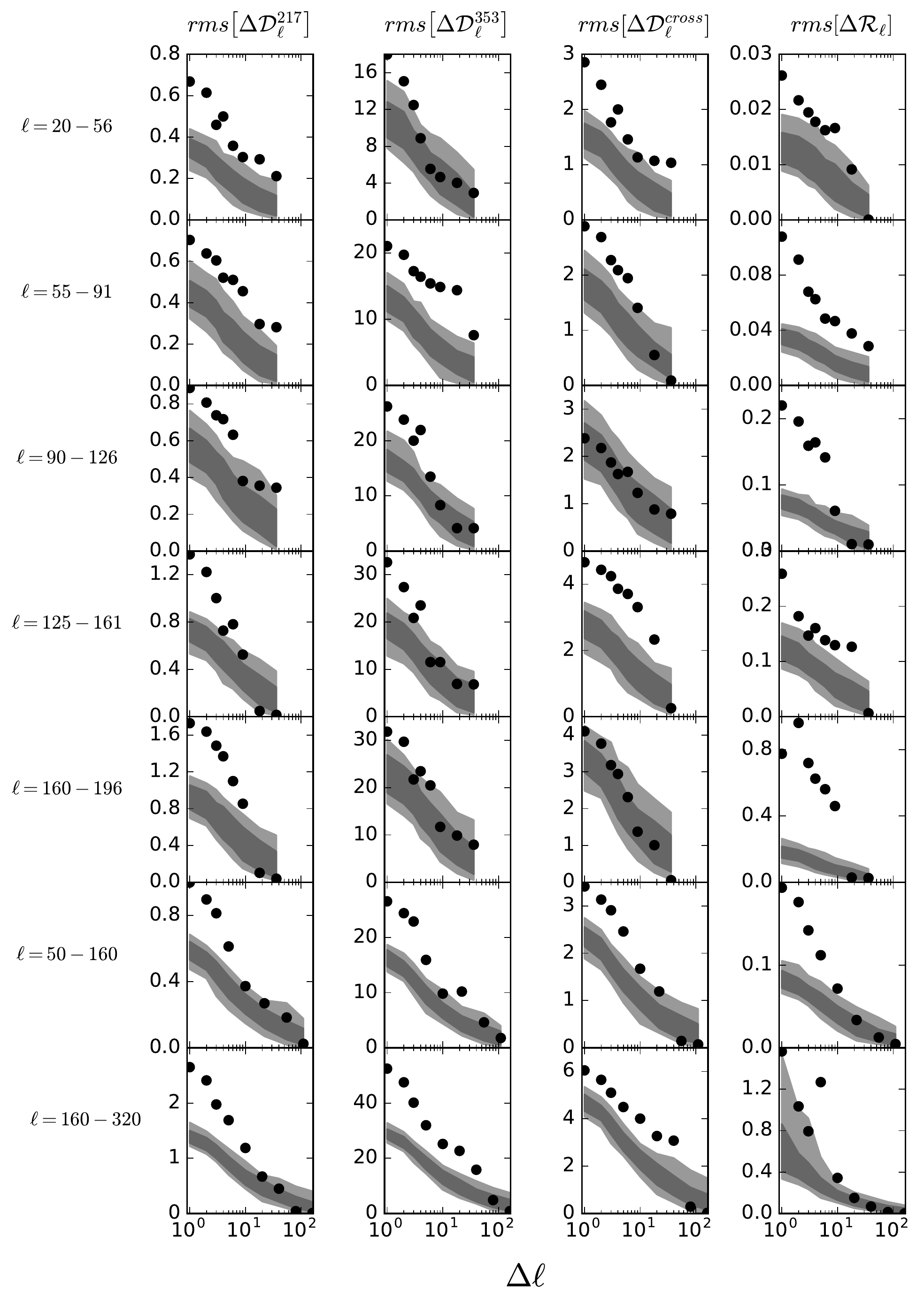}
    \end{tabular}
  \end{center}
  \caption[example] { \label{fig:syst2} Root-mean-square (rms) of the difference
    $\Delta\dl^{BB}$ and $\Delta\rl^{BB}$ between the two independent cross
    spectra of the four HM$_i$DS$_j$ data splits. The left three columns show
    $\Delta\dl^{BB}$ in $\muK^2$ for $217\times217$, $353\times353$, and
    $217\times353$. The right column shows $\Delta\rl^{BB}$.  Each row shows the
    rms computed in the indicated broad multipole bin. The $x$-axis of each
    panel indicates the fine binning of the raw spectra from which the rms
    values are calculated in the broad bins. The binning ranges from
    $\Delta\ell=1$ to the full bin width. The dark and light gray regions
    enclose $68\%$ and $95\%$ of the corresponding signal+noise simulations.}
\end{figure}

Figure~\ref{fig:syst4} shows the bias on $\rl^{BB}$ in each LR region from a
correlated systematic of the magnitude of the uncorrelated systematic measured
in the LR72 region. We note again that the excess power in the HM map difference
null test does not appear to change in amplitude in smaller sky fractions.

To assess the level of correlated systematics in the HM and DS splits, which
will not contaminate the map difference null tests, we
perform a difference-of-bandpowers null test on $\dl^{BB}$ and $\rl^{BB}$
computed from the HM$_i$DS$_j$ splits. Each of these four splits contains 1/4th
of the total nominal mission data and is independent of the others.  (For
example, HM$_1$DS$_2$ is the quarter of the data that belongs to half-mission
one and detector-set two.)  We compute bandpowers and $\rl^{BB}$ from what
should be the two maximally uncontaminated cross spectra: HM$_1$DS$_2$ $\times$
HM$_2$DS$_1$ and HM$_1$DS$_1$ $\times$ HM$_2$DS$_2$.  We then take the
difference of bandpowers, $\Delta\dl^{BB}$, and the difference of the
correlation ratios, $\Delta\rl^{BB}$ and compare them to the corresponding
differences calculated from signal+noise simulations. The simulations are
constructed from a fixed signal realization and 100 noise realizations
constructed from the four $QU$ covariance maps of the HM$_i$DS$_j$ splits.

The top and middle panels of Figure~\ref{fig:syst1} show the two independently
measured $\rl^{BB}$ in bins of $\Delta\ell=3$ along with the 100 signal+noise
simulations for reference. The bottom panel shows the difference,
$\Delta\rl^{BB}$ . (The simulation realizations are omitted in the bottom panel
for clarity but, similarly to the top and middle panels, they show no outlying
realizations.) Deviation of $\Delta\rl^{BB}$ from zero in the bottom panel is
evidence for correlated systematic contamination between the HM and DS split
halves. The outlier points in the top and bottom panel are real. However, only
100 simulation realizations are plotted, and because of the high side tail of
the likelihood distribution ($\rl^{BB}$ is a ratio whose denominator can be
close to zero due to noise fluctuations) the likelihood of these fluctuations is
probably underestimated by the size of the error bars. 
The known uncorrelated systematics act to increase the effective noise
in this null test, which we do not account for. Nevertheless, the excess uncorrelated
power is $\lesssim20\%$ of the total noise power and thus cannot explain
the observed discrepancies.

\begin{figure*}[]
  \begin{center}
    \begin{tabular}{c}
      \includegraphics[width=2\columnwidth]{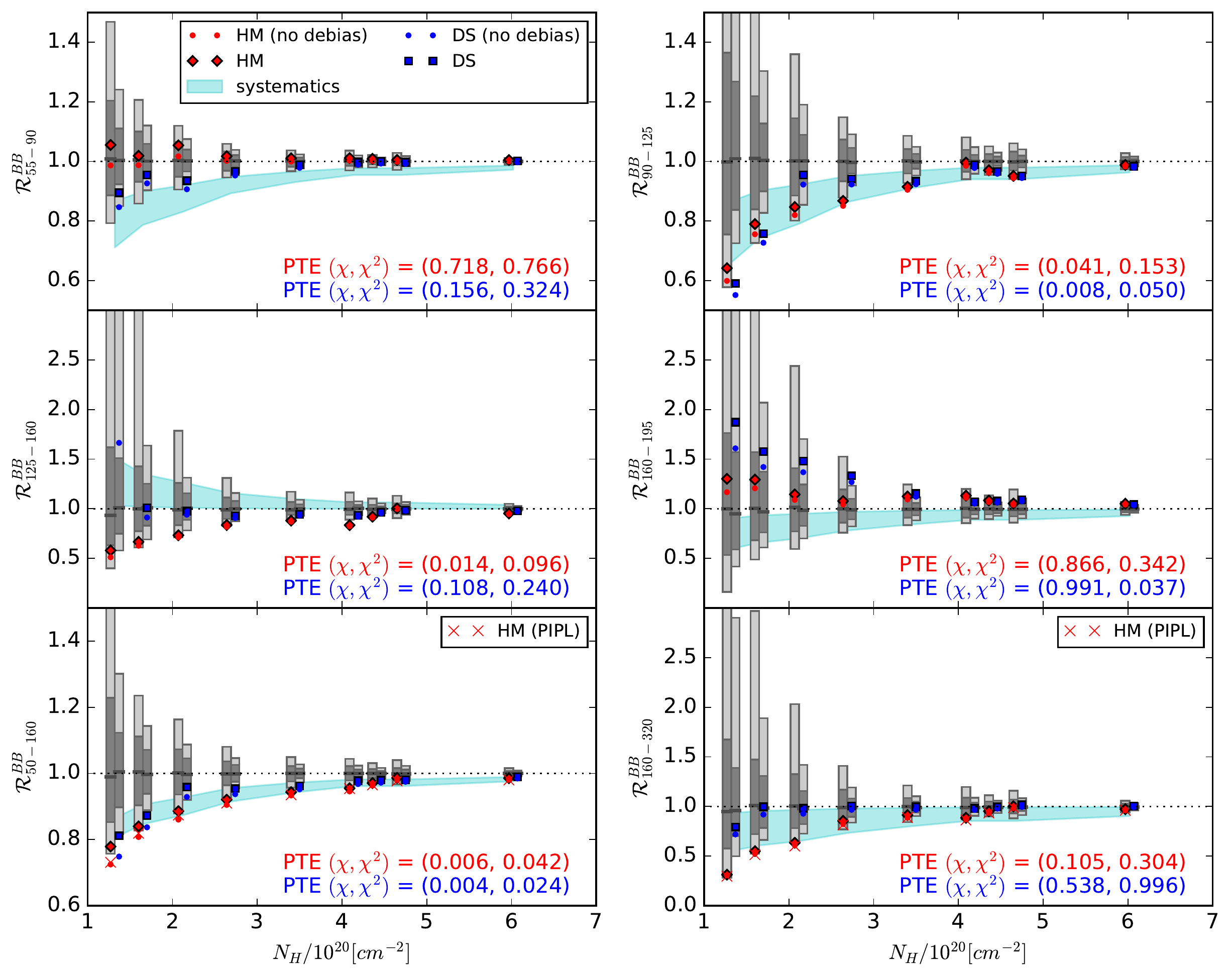}
    \end{tabular}
  \end{center}
  \caption[example] { \label{fig:nh_curve} $\rl^{BB}$ in the 9 LR regions,
    plotted as a function of neutral hydrogen column density, $\nHI$. Each panel
    is a different multipole bin. The red diamonds and and blue squares show
    $\rl^{BB}$ calculated from the HM and DS splits, respectively. The
    corresponding red and blue dots show $\rl^{BB}$ without accounting for noise
    bias. The vertical bars indicate the regions enclosing $68\%$ and $95\%$ of
    the signal+noise simulations ($16 - 84$ and $2.5-97.5$ percentiles,
    respectively) and the dark gray horizontal lines show the median of the
    simulations. The red x's in the bottom two panels show the corresponding
    PIPL points, taken from the appendix, which should be the same as the ``HM
    (no debias)'' points. In each panel there are four statistics listed: the
    $\chi$ and $\chi^2$ PTE for the HM and DS splits, calculated as the number
    of simulations having $\chi$ ($\chi^2$) less (greater) than the observed
    value. A low/high $\chi$ PTE indicates data that is coherently low/high. A 
    low/high $\chi^2$ PTE indicates data that has too much/little scatter. The
    PTEs do not account for systematic uncertainty. The cyan shaded region
    indicates the region between the pessimistic and optimistic estimates of the
    systematic bias on $\rl^{BB}$, computed as the expectation for
    $\rl^{BB}$ in the presence of a systematic upward bias on $\dl^{BB}$ given
    by Eq.~\ref{eq:dlsyst} and the $\Delta\ell=3$ (pessimistic) and
    $\Delta\ell=$~full width/2 (optimistic) data in Figure~\ref{fig:syst2}.}
\end{figure*}

\begin{figure}[]
  \begin{center}
    \begin{tabular}{c}
      \includegraphics[width=1\columnwidth]{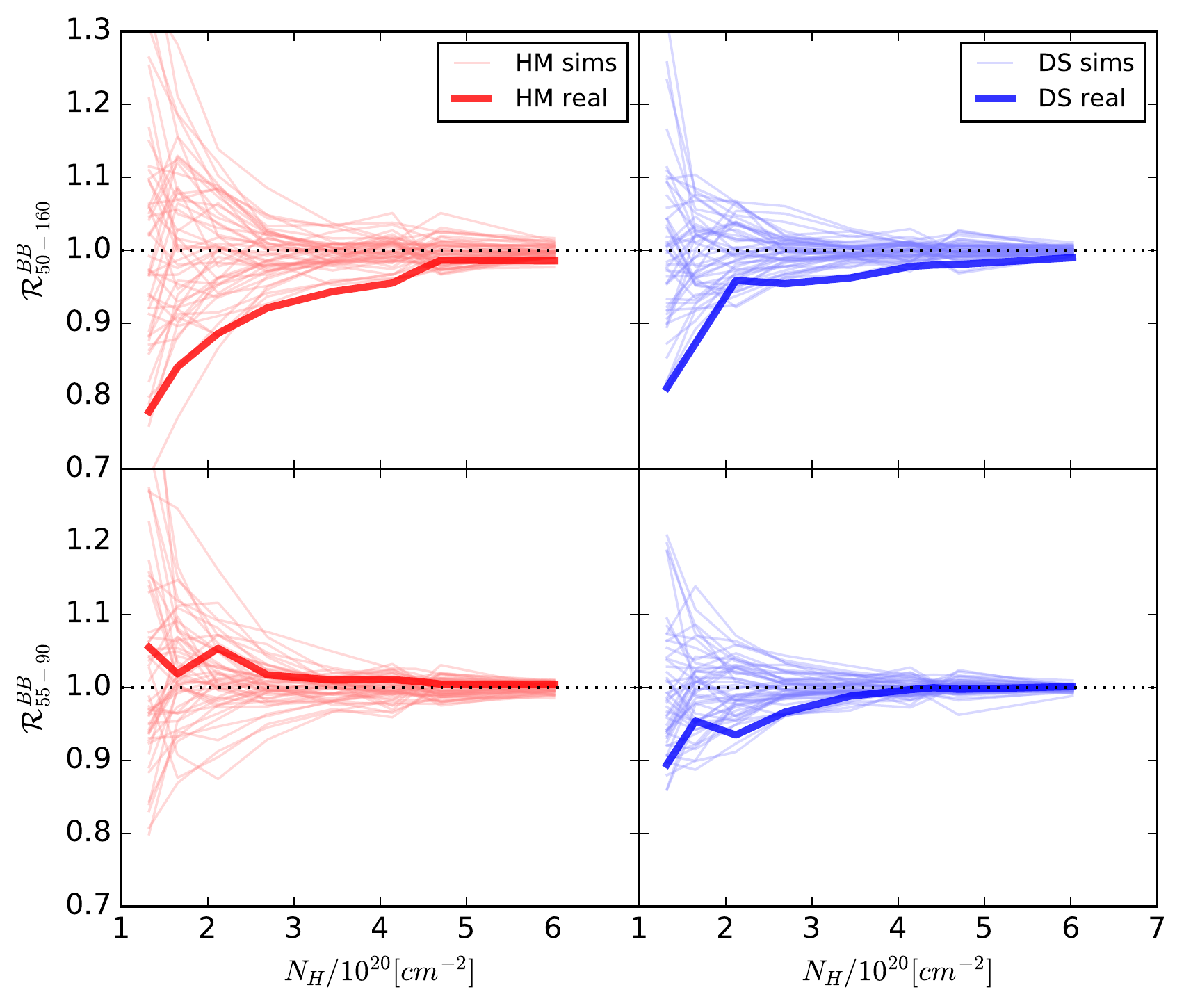}
    \end{tabular}
  \end{center}
  \caption[example] { \label{fig:nh_sims} $\rl^{BB}$ in the 9 LR regions 
    plotted as a function of neutral hydrogen column density, $\nHI$. The left
    and right columns are the HM and DS splits, respectively. The top and
    bottom rows are the $\ell=50-160$ bin and $\ell=55-90$ bin, respectively.
    The thick line shows the real data and is the same as the noise debiased
    points in the corresponding panels of Figure~\ref{fig:nh_curve}. The thin
    lines show the first 50 signal+noise realizations. The only component held
    fixed between LR regions in a single simulation realization is the $QU$
    covariance generated noise map.}
\end{figure}

Figure~\ref{fig:syst2} shows the root-mean-square (rms) of
$\Delta\rl^{BB}$ and $\Delta\dl^{BB}$ (not shown) calculated in 7 different
$\ell$ bins. Also plotted are the regions enclosing $68\%$ and $95\%$ of the rms
values from simulation. The rms is calculated in each $\ell$ bin from more
finely binned data with fine bin width $\Delta\ell$ ranging from 1 to the full
bin width, purposefully chosen in some bins to be $\Delta \ell=36$, which has many
divisors, instead of $\Delta\ell=35$, which does not. (For instance, 
Figure~\ref{fig:syst1} shows $\Delta\rl^{BB}$ in bins
of $\Delta\ell=3$, and the corresponding $rms[\rl^{BB}]$ is plotted as the
$\Delta\ell=3$ points in the right hand column of Figure~\ref{fig:syst2}.)  When
$\Delta\ell$ is equal to the full bin width, the rms is simply the absolute
value of the difference of two points.

We find strong disagreement between the observed and simulated rms values in
finely binned spectra. For the full bin width, we find general agreement, except
for in the $\ell=20-56$ and $\ell=55-91$ bin. The behavior
of the observed rms indicates a systematic that at least partially averages down
when binning in $\ell$. We can see this in Figure~\ref{fig:syst1}: apparent
correlated structure in $\Delta\rl^{BB}$ will average to zero in broad $\ell$
bins, resulting in agreement with simulations. It is unknown whether the total
correlated systematics will also average down -- the bandpower difference null
test uncovers systematics that are correlated within each of the two pairs of
HM$_i$DS$_j$ and yet produce a different bias in the two cross
spectra. Systematics that produce the same bias in each cross spectrum will not
produce null test failures.

We therefore adopt $\Delta\ell=\mathrm{full~width}/2$ as an optimistic
estimate of the systematic contamination due to instrumental systematics and 
$\Delta\ell=3$ as a pessimistic case. We then compute the expected bias on $\dl$
as one half the observed minus the mean simulated rms:

\begin{equation}\label{eq:dlsyst}
\dl^{s,syst} = \frac{1}{2}\left(rms\left[\Delta\dl^{s,obs}\right] -
\left<rms\left[\Delta\dl^{s,sim}\right]\right> \right)
\end{equation}

\noindent where $<>$ is the mean over simulations; $s = 217$, $353$, or
$cross = 217\times353$; and the factor $1/2$ assumes that the magnitude of the
null test failure is twice the contamination in each cross spectrum
individually. We set $\dl^{syst}=0$ if it is $<0$. We then use these
estimates of the bias in the next section to estimate the bias on $\rl^{BB}$ in
each LR region.

\section{Significance of measurements}
\label{sec:skyfrac}

\subsection{Trends with sky fraction}

Figure~\ref{fig:nh_curve} shows $\rl^{BB}$ in 6 separate $\ell$ bins, plotted as
a function of the mean neutral hydrogen column density, $\nHI$, in each LR
region as reported in PIPL. The HM data in the bottom two panels are directly comparable to
$\rl^{BB}$ reported in PIPL, and we find very good agreement. (We note that this
agreement is true of the $\rl^{BB}$ values shown in the histograms plotted in
the appendix of PIPL. The
$\mathcal{R}^{BB}_{50-160}$ points plotted in Figure~3 of PIPL appear inconsistent
with both the current results and the appendix of PIPL.)  The HM and DS splits
are both plotted. Vertical bars indicate the regions enclosing $68\%$ and $95\%$
of the signal+noise simulations.  Lastly, we plot the region between the ``optimistic'' and
``pessimistic'' systematic bias predictions discussed in the previous section
and defined in Eq.~\ref{eq:dlsyst}.

To gauge the significance of any trends in Figure~\ref{fig:nh_curve}, we compute
two statistics from both the real data and each simulation realization: the inverse
variance weighted $\chi$ and $\chi^2$, defined as

\begin{equation}
  \chi^n = \sum_{\mathrm{LR}} \left[ (\rl^{BB} - 1)/\sigma_{\rl^{BB}} \right]^n .
\end{equation}

\noindent where $\sigma_{\rl^{BB}}$ is the width of the $68\%$ confidence
intervals shown in Figure~\ref{fig:nh_curve}. We then compute the probability to
exceed (PTE) of these statistics. The
$\chi$ PTE is defined as the fraction of simulations having $\chi$ \textit{less}
than the observed value, so that low/high PTEs indicate $\rl^{BB}$ which is
coherently low/high across LR regions. The $\chi^2$ PTE is defined as
the fraction of simulations having $\chi^2$ \textit{greater} than the observed
value, so that a low PTE indicates data with too much scatter under the no
decorrelation hypothesis.

Without accounting for instrumental systematics, the strongest disagreement with
simulations comes from the $\ell=50-160$ DS split, with PTE$_{\chi}=0.4\%$. The
HM and DS split appear qualitatively consistent in this bin. Examining the
$\ell=50-160$ sub-bins, however, we find, different results. Neither the
$\ell=55-90$ bin, which has similar signal-to-noise to the full $\ell=50-160$
bin, nor the $\ell=125-160$ bin show strong evidence for decorrelation. In these two
bins, apparent trends in either the DS or HM splits are not seen in the
other split. The $\ell=90-125$ bin shows the largest downward deviation of
$\rl^{BB}$ from $1$ and has qualitative consistency between HM and DS. 

The marginally low $\rl^{BB}$ in the $\ell=50-160$ and $\ell=90-125$ bins are,
however, fully consistent with the estimate of bias from instrumental
systematics. Furthermore, the $\ell=160-195$ bin has DS PTE$_{\chi}=0.991$,
which is marginal evidence for the unphysical $\rl^{BB}>1$. There is also an
apparent trend to higher $\rl^{BB}$ with lower $\nHI$ in this bin. This upward
bias is possible if systematics correlate between 217 and 353, a possibility
Figure~\ref{fig:syst2} shows some evidence for. The ``optimistic'' systematics
line in the $\ell=125-160$ panel of Figure~\ref{fig:nh_curve} shows a positive
bias because the corresponding $\Delta\ell=18$ $rms[\dl]$ values in Figure~\ref{fig:syst2} show
no disagreement with simulations in $217\times217$ or $353\times353$ but a
significant positive bias in $217\times353$. This bin does show an upward fluctuation of $\rl^{BB}$ in LR16
prior to noise debiasing. (After noise debiasing, $217\times217$ becomes
negative and $\rl^{BB}$ becomes undefined.) We note, however, that in the $\ell=160-195$
bin, Figure~\ref{fig:syst2} shows no evidence for problems
with the $217\times353$ cross spectrum.

Figure~\ref{fig:nh_sims} shows the $\mathcal{R}_{50-160}^{BB}$ and
$\mathcal{R}_{55-90}^{BB}$ data from Figure~\ref{fig:nh_curve} plotted as thick
lines and the first 50 realizations of the signal+noise simulations plotted as
thin lines. Clear trends are visible in the simulations indicating significant
correlation between LR regions. The PTEs listed in Figure~\ref{fig:nh_curve}
would be much more significant were it not for these correlations. The
simulations are the $QU$ covariance noise realizations plus a Gaussian
dust + CMB signal realization. Only the noise realization is
common between the LR regions in a given realization. As in PIPL, the dust and CMB
realizations change. We therefore conclude that the noise common to the nested
LR regions is responsible for the correlations. This result is perhaps
unsurprising given that $\rl^{BB}$ is measured largely without sample
variance. In simulations substituting the fixed PySM dust + CMB realization for
the varying Gaussian dust + CMB realizations, we observe nearly identical
correlations between LR regions. We also observe nearly identical correlations
when we substitute in the FFP8 noise simulations for the $QU$ covariance noise
realizations. 

\begin{figure}[]
  \begin{center}
    \begin{tabular}{c}
      \includegraphics[width=1\columnwidth]{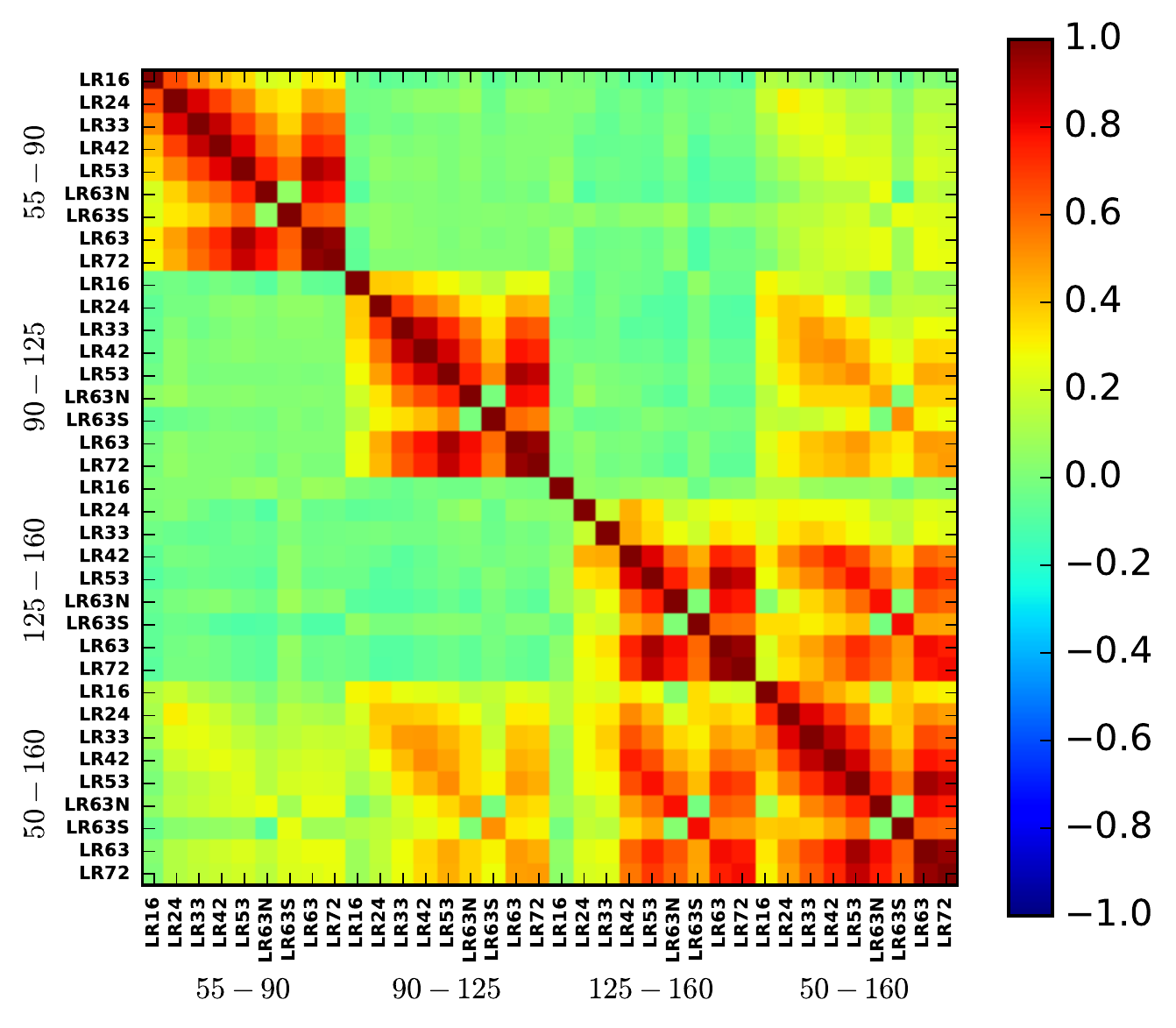}
    \end{tabular}
  \end{center}
  \caption[example] { \label{fig:corr_matrix} $\rl^{BB}$ correlation coefficient
    between multipole bins and LR regions.}
\end{figure}

To see the correlations more clearly, Figure~\ref{fig:corr_matrix} shows the
correlation coefficient matrix for $\rl^{BB}$ between multipole bins and LR
regions. In each bin, there are large correlations between LR regions except for
63N and 63S, which are non-overlapping. These correlations ensure that strong
trends in $\rl^{BB}$ as a function of $\nHI$ are expected even with no
decorrelation. Measurements of $\rl^{BB}$ in different LR regions may therefore
not be regarded as approximately statistically independent, as advocated in
PIPL. We do note that non-overlapping $\ell$ bins appear to be negligibly
correlated, as expected.

\begin{figure}[]
  \begin{center}
    \begin{tabular}{c}
      \includegraphics[width=1\columnwidth]{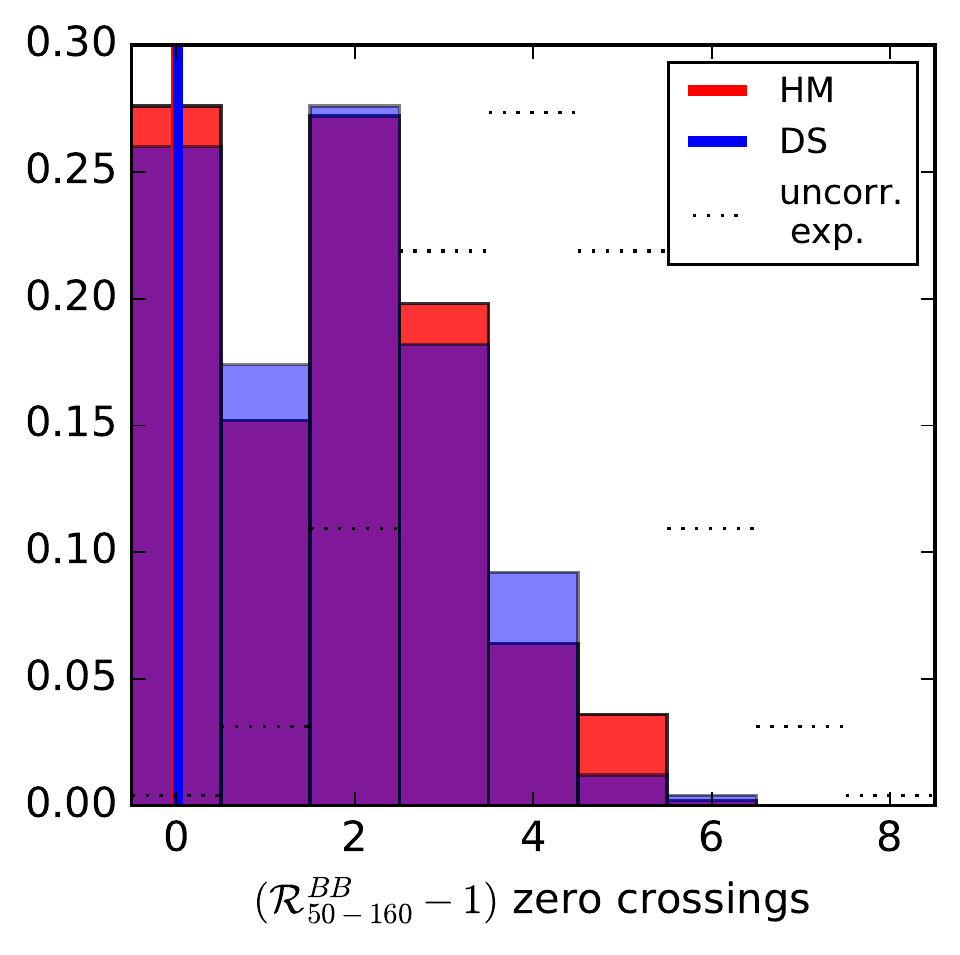}
    \end{tabular}
  \end{center}
  \caption[example] { \label{fig:nh_zerocross} Number of zero crossings of
    $[\mathcal{R}^{BB}_{50-160}(\nHI)-1]$ calculated from HM and DS splits (red
    and blue) and shown in the top left panel of Figure~\ref{fig:nh_curve}. The
    vertical lines are the observed values and the histograms are the
    distribution from the signal+noise simulations. The dotted histogram is the
    expectation for 9 uncorrelated random numbers distributed about 1 computed
    from the binomial distribution.}
\end{figure}

We also compute the number of zero crossings of $\rl^{BB}(\nHI)-1$ in the
$\ell=50-160$ bin, which we show in Figure~\ref{fig:nh_zerocross}. The HM 
and DS data both have no zero crossings.  If measurements of
$\rl^{BB}$ in different LR regions were uncorrelated, these results would be
highly unlikely without significant dust decorrelation. This is evident from the
dotted line histogram, which is the number of zero crossings predicted by the
binomial distribution under the hypothesis that $\rl>1$ and $\rl<1$ are equally
likely. We find, however, that in the simulations, which account for correlations
between LR regions, observing no zero crossings is in fact one of the most
likely outcomes.

\subsection{Maximum Likelihood $\rl^{BB}$}

Table~\ref{tab:ML} lists the maximum likelihood (ML) values of the noise
debiased $\rl^{BB}$ in each LR region and in different $\ell$ bins. (We
also list the $\ell=50-160$ non-noise-debiased values for comparison.)  We quote
statistical uncertainties as empirically determined  
from the simulations. We adjust the simulations by adding or subtracting a constant value to
each realization's $217\times353$ binned $\dl$'s such that the mean
decorrelation of the signal-only simulations equals the observed value. We
leave the $217\times217$ and $353\times353$ $\dl$'s alone under the assumption
that small levels of decorrelation suppress power in the cross spectrum without
significantly affecting the auto spectra.  We then recompute $\rl^{BB}$ of each
realization and adopt this as the ML distribution. The statistical
uncertainties, $\sigma_{\rl^{BB}}$, quoted in Table~\ref{tab:ML} are defined
such that $\rl^{BB} \pm 2\sigma_{\rl^{BB}}$ encloses $95\%$ of the adjusted
simulations ($2.5 - 97.5$ percentiles). We also include data for the
$\ell=20-55$ bin in LR72 and LR63. The dust amplitude is strong enough in these
region that the known low-$\ell$ \planck\ systematics appear to produce only a
small bias on $\rl$. The systematic uncertainty quoted is the mean of the
systematics region shown in Figure~\ref{fig:nh_curve}.

Because of the significant systematic uncertainty in every bin and every LR
region, we advocate that the ML values listed in Table~\ref{tab:ML} only be
interpreted in light of the systematic uncertainty. The absence of evidence for
decorrelation in the $\ell=55-90$ bin therefore places strong constraints on the
maximum possible level of decorrelation at the peak of the expected inflationary
$B$-mode signal. For instance, taken at face value, dust model 2 of PySM is
consistent with $\mathcal{R}_{90-125}^{BB}$ and $\mathcal{R}_{125-160}^{BB}$
measured in LR63 (see Figure~\ref{fig:Rfine_LR63}), but the model appears strongly ruled
out by $\mathcal{R}_{55-90}^{BB}$. Also apparently inconsistent is the flat decorrelation of
$\mathcal{R}^{BB}_{50-160}=0.95$ assumed by PIPL to predict an expected bias 
on $r$.

Table~\ref{tab:PTE} lists the corresponding $\rl^{BB}$ PTE values, defined as
the fraction of simulations having $\rl^{BB}$ less than the observed value. 
The PTEs do not account for systematic uncertainty.

\section{Conclusions}
\label{sec:conclusions}

In this paper we have revisited the the evidence for decorrelation in
the polarized  dust signal in \planck\ data. We have made several
improvements in our analysis over the \planck\ analysis. Our
conclusions can be summarized as follows:

\begin{itemize}
\item The destriping procedure correlates noise between data splits, a small but
  statistically relevant bias that cross-correlation power spectrum estimation
  must correct for to avoid artificially lowering $\rl$ measurements. 

\item The data split difference maps contain excess power that is not present in
  the FFP8 simulations, thus indicating the presence of uncorrelated
  systematics. We find that if contamination were present at this level in
  the cross spectra it would push $\rl$ measurements far below the observed
  values. By using quarter data splits, we have estimated the order of magnitude
  of correlated systematics, which will bias $\rl$. Since we find evidence for
  these systematics and cannot exclude that they will average down to negligible
  levels in broad bins, we conclude that $\rl$ measurements should
  only be interpreted in light of the systematic uncertainties shown in
  Figure~\ref{fig:nh_curve} and quoted in Table~\ref{tab:ML}.

\item Even taking the $\rl$ measurements at face value, at a fixed angular scale,
  the results from nested sky cuts are heavily correlated. Once these
  correlations are taken into account, the evidence for deviation from unity
  weakens significantly.

\end{itemize}

We have employed two statistics to quantify the discrepancy from the null
hypothesis $\rl^{BB}=1$ everywhere. The $\chi^2$ statistic calculates the
average discrepancy with unity correlation while the $\chi$ statistic measures
coherent shifts upwards or downwards. Although both statistics are
generated using diagonal errors (of very strongly correlated covariance matrix),
they are compared to simulations so that PTE values are valid (in the
absence of systematics). Statistical evidence in the absence of systematics is weak,
$2-3$ sigma. However, since we demonstrate the presence of an unknown systematic
that can affect results at the level of the measurement accuracy, we must
conservatively conclude that there is no statistically compelling evidence for
decorrelation in the \planck\ data. Additional multifrequency data will be
required to place stronger constraints on decorrelation.

\section*{Acknowledgments}

The authors would like to thank Julian Borrill for making the \planck\ FFP8
noise realizations available and for answering our many questions about them. We
thank Tuhin Ghosh for useful discussions.

\bibliography{ms}

\appendix

\newpage
\begin{turnpage}
\begin{table*}[tbp!] 
\centering 
\caption{Noise debiased $\rl^{BB}$ in different multipole bins and
        different LR regions measured with the HM and DS splits. The undebiased
        $\ell=50-160$ bin is also listed for comparison to the noise debiased
        values. The quoted statistical uncertainties are one half of the region
        enclosing $95\%$ of the adjusted signal+noise simulations. The
        systematic uncertainties are the mean of the optimistic and pessimistic systematics estimates
        shown in Figure~\ref{fig:nh_curve}.} 
\label{tab:ML} 
\begin{tabular}{lcccccccccc} 
\addlinespace[1ex] 
\hline 
\hline 
\addlinespace[1ex] 
\rule{0pt}{2ex} 
  & \mc{LR16}& \mc{LR24}& \mc{LR33}& \mc{LR42}& \mc{LR53}& \mc{LR63N}& \mc{LR63}& \mc{LR63S}& \mc{LR72} \\ 
\addlinespace[1ex] 
\hline 
\addlinespace[1ex] 
  $f_{\rm sky}^{\rm eff}$ [\%]\hfil& 16& 24& 33& 42& 53& 33& 63& 30& 72\\ 
\addlinespace[1ex] 
\hline 
\hline 
\addlinespace[1ex] 
\addlinespace[1ex] 
$\ell$ range & \multicolumn{9}{c}{Maximum Likelihood $\rl^{BB} (ML\substack{ +stat./syst.\\ -stat./syst.})$ \bigg( \begin{tabular}{@{}c@{}} HM \\[.1cm] DS \end{tabular} \bigg)   } \\\addlinespace[1ex] 
\addlinespace[1ex] 
\hline 
\addlinespace[1ex] 
\addlinespace[1ex] 
 \begin{tabular}{@{}c@{}}50--160\\ (no d.b.) \end{tabular} & \begin{tabular}{@{}c@{}}$0.725\substack{ +0.20/0.16\\ -0.10/0.00 }$ \\[.2cm]$0.748\substack{ +0.12/0.16\\ -0.08/0.00 }$\end{tabular}& \begin{tabular}{@{}c@{}}$0.808\substack{ +0.10/0.12\\ -0.07/0.00 }$ \\[.2cm]$0.837\substack{ +.064/.124 \\ -.048/.000 }$\end{tabular}& \begin{tabular}{@{}c@{}}$0.860\substack{ +.075/.100 \\ -.053/.000 }$ \\[.2cm]$0.929\substack{ +.043/.100 \\ -.036/.000 }$\end{tabular}& \begin{tabular}{@{}c@{}}$0.905\substack{ +.038/.065 \\ -.033/.000 }$ \\[.2cm]$0.937\substack{ +.023/.065 \\ -.020/.000 }$\end{tabular}& \begin{tabular}{@{}c@{}}$0.933\substack{ +.024/.043 \\ -.021/.000 }$ \\[.2cm]$0.952\substack{ +.014/.043 \\ -.013/.000 }$\end{tabular}& \begin{tabular}{@{}c@{}}$0.945\substack{ +.022/.028 \\ -.018/.000 }$ \\[.2cm]$0.968\substack{ +.012/.028 \\ -.012/.000 }$\end{tabular}& \begin{tabular}{@{}c@{}}$0.963\substack{ +.015/.028 \\ -.012/.000 }$ \\[.2cm]$0.972\substack{ +.008/.028 \\ -.008/.000 }$\end{tabular}& \begin{tabular}{@{}c@{}}$0.980\substack{ +.020/.028 \\ -.016/.000 }$ \\[.2cm]$0.974\substack{ +.011/.028 \\ -.010/.000 }$\end{tabular}& \begin{tabular}{@{}c@{}}$0.980\substack{ +.008/.018 \\ -.007/.000 }$ \\[.2cm]$0.985\substack{ +.004/.018 \\ -.004/.000 }$\end{tabular}\\[1cm] 
\addlinespace[1ex] 
\addlinespace[1ex] 
 50--160 & \begin{tabular}{@{}c@{}}$0.779\substack{ +0.21/0.16\\ -0.10/0.00 }$ \\[.2cm]$0.811\substack{ +0.13/0.16\\ -0.08/0.00 }$\end{tabular}& \begin{tabular}{@{}c@{}}$0.840\substack{ +0.11/0.12\\ -0.07/0.00 }$ \\[.2cm]$0.872\substack{ +.066/.124 \\ -.049/.000 }$\end{tabular}& \begin{tabular}{@{}c@{}}$0.886\substack{ +.077/.100 \\ -.054/.000 }$ \\[.2cm]$0.958\substack{ +.044/.100 \\ -.037/.000 }$\end{tabular}& \begin{tabular}{@{}c@{}}$0.921\substack{ +.039/.065 \\ -.033/.000 }$ \\[.2cm]$0.954\substack{ +.024/.065 \\ -.021/.000 }$\end{tabular}& \begin{tabular}{@{}c@{}}$0.943\substack{ +.024/.043 \\ -.021/.000 }$ \\[.2cm]$0.962\substack{ +.014/.043 \\ -.013/.000 }$\end{tabular}& \begin{tabular}{@{}c@{}}$0.955\substack{ +.022/.028 \\ -.018/.000 }$ \\[.2cm]$0.978\substack{ +.012/.028 \\ -.012/.000 }$\end{tabular}& \begin{tabular}{@{}c@{}}$0.971\substack{ +.015/.028 \\ -.012/.000 }$ \\[.2cm]$0.980\substack{ +.008/.028 \\ -.008/.000 }$\end{tabular}& \begin{tabular}{@{}c@{}}$0.986\substack{ +.020/.028 \\ -.016/.000 }$ \\[.2cm]$0.980\substack{ +.011/.028 \\ -.010/.000 }$\end{tabular}& \begin{tabular}{@{}c@{}}$0.985\substack{ +.008/.018 \\ -.007/.000 }$ \\[.2cm]$0.990\substack{ +.004/.018 \\ -.004/.000 }$\end{tabular}\\[1cm] 
\addlinespace[1ex] 
\addlinespace[1ex] 
 20--55 & \ldots& \ldots& \ldots& \ldots& \ldots& \begin{tabular}{@{}c@{}}$1.001\substack{ +.004/.008 \\ -.004/.000 }$ \\[.2cm]$1.005\substack{ +.003/.008 \\ -.003/.000 }$\end{tabular}& \begin{tabular}{@{}c@{}}$0.994\substack{ +.003/.008 \\ -.003/.000 }$ \\[.2cm]$1.003\substack{ +.002/.008 \\ -.002/.000 }$\end{tabular}& \begin{tabular}{@{}c@{}}$0.987\substack{ +.004/.008 \\ -.004/.000 }$ \\[.2cm]$1.001\substack{ +.003/.008 \\ -.003/.000 }$\end{tabular}& \begin{tabular}{@{}c@{}}$0.996\substack{ +.002/.005 \\ -.002/.000 }$ \\[.2cm]$1.001\substack{ +.001/.005 \\ -.001/.000 }$\end{tabular}\\[1cm] 
\addlinespace[1ex] 
\addlinespace[1ex] 
 55--90 & \begin{tabular}{@{}c@{}}$1.055\substack{ +0.25/0.21\\ -0.11/0.00 }$ \\[.2cm]$0.895\substack{ +0.11/0.21\\ -0.07/0.00 }$\end{tabular}& \begin{tabular}{@{}c@{}}$1.019\substack{ +0.11/0.16\\ -0.07/0.00 }$ \\[.2cm]$0.954\substack{ +.059/.157 \\ -.047/.000 }$\end{tabular}& \begin{tabular}{@{}c@{}}$1.054\substack{ +.063/.125 \\ -.049/.000 }$ \\[.2cm]$0.935\substack{ +.037/.125 \\ -.030/.000 }$\end{tabular}& \begin{tabular}{@{}c@{}}$1.017\substack{ +.031/.080 \\ -.027/.000 }$ \\[.2cm]$0.967\substack{ +.019/.080 \\ -.019/.000 }$\end{tabular}& \begin{tabular}{@{}c@{}}$1.010\substack{ +.019/.052 \\ -.016/.000 }$ \\[.2cm]$0.988\substack{ +.012/.052 \\ -.011/.000 }$\end{tabular}& \begin{tabular}{@{}c@{}}$1.011\substack{ +.018/.034 \\ -.015/.000 }$ \\[.2cm]$0.997\substack{ +.011/.034 \\ -.009/.000 }$\end{tabular}& \begin{tabular}{@{}c@{}}$1.009\substack{ +.011/.034 \\ -.010/.000 }$ \\[.2cm]$1.000\substack{ +.006/.034 \\ -.007/.000 }$\end{tabular}& \begin{tabular}{@{}c@{}}$1.005\substack{ +.015/.034 \\ -.014/.000 }$ \\[.2cm]$0.998\substack{ +.009/.034 \\ -.008/.000 }$\end{tabular}& \begin{tabular}{@{}c@{}}$1.005\substack{ +.006/.021 \\ -.005/.000 }$ \\[.2cm]$1.001\substack{ +.004/.021 \\ -.004/.000 }$\end{tabular}\\[1cm] 
\addlinespace[1ex] 
\addlinespace[1ex] 
 90--125 & \begin{tabular}{@{}c@{}}$0.641\substack{ +0.39/0.24\\ -0.19/0.00 }$ \\[.2cm]$0.590\substack{ +0.26/0.24\\ -0.11/0.00 }$\end{tabular}& \begin{tabular}{@{}c@{}}$0.789\substack{ +0.28/0.18\\ -0.12/0.00 }$ \\[.2cm]$0.757\substack{ +0.12/0.18\\ -0.07/0.00 }$\end{tabular}& \begin{tabular}{@{}c@{}}$0.847\substack{ +0.15/0.14\\ -0.09/0.00 }$ \\[.2cm]$0.954\substack{ +.093/.144 \\ -.068/.000 }$\end{tabular}& \begin{tabular}{@{}c@{}}$0.868\substack{ +.064/.093 \\ -.055/.000 }$ \\[.2cm]$0.941\substack{ +.045/.093 \\ -.039/.000 }$\end{tabular}& \begin{tabular}{@{}c@{}}$0.915\substack{ +.041/.061 \\ -.034/.000 }$ \\[.2cm]$0.934\substack{ +.024/.061 \\ -.022/.000 }$\end{tabular}& \begin{tabular}{@{}c@{}}$0.996\substack{ +.041/.040 \\ -.029/.000 }$ \\[.2cm]$0.988\substack{ +.024/.040 \\ -.020/.000 }$\end{tabular}& \begin{tabular}{@{}c@{}}$0.970\substack{ +.025/.040 \\ -.022/.000 }$ \\[.2cm]$0.967\substack{ +.015/.040 \\ -.015/.000 }$\end{tabular}& \begin{tabular}{@{}c@{}}$0.951\substack{ +.030/.040 \\ -.028/.000 }$ \\[.2cm]$0.951\substack{ +.020/.040 \\ -.019/.000 }$\end{tabular}& \begin{tabular}{@{}c@{}}$0.988\substack{ +.014/.025 \\ -.013/.000 }$ \\[.2cm]$0.984\substack{ +.008/.025 \\ -.008/.000 }$\end{tabular}\\[1cm] 
\addlinespace[1ex] 
\addlinespace[1ex] 
 125--160 & \begin{tabular}{@{}c@{}}$0.580\substack{ +0.83/0.00\\ -0.27/-0.27 }$ \\[.2cm]\ldots\end{tabular}& \begin{tabular}{@{}c@{}}$0.666\substack{ +0.67/0.00\\ -0.17/-0.18 }$ \\[.2cm]$1.009\substack{ +0.33/0.00\\ -0.15/-0.18 }$\end{tabular}& \begin{tabular}{@{}c@{}}$0.732\substack{ +0.32/0.00\\ -0.12/-0.14 }$ \\[.2cm]$0.976\substack{ +0.16/0.00\\ -0.10/-0.14 }$\end{tabular}& \begin{tabular}{@{}c@{}}$0.836\substack{ +0.14/0.00\\ -0.08/-0.08 }$ \\[.2cm]$0.929\substack{ +.077/.000 \\ -.058/-.080 }$\end{tabular}& \begin{tabular}{@{}c@{}}$0.880\substack{ +.078/.000 \\ -.053/-.049 }$ \\[.2cm]$0.944\substack{ +.045/.000 \\ -.034/-.049 }$\end{tabular}& \begin{tabular}{@{}c@{}}$0.835\substack{ +.072/.000 \\ -.048/-.031 }$ \\[.2cm]$0.934\substack{ +.036/.000 \\ -.029/-.031 }$\end{tabular}& \begin{tabular}{@{}c@{}}$0.919\substack{ +.048/.000 \\ -.034/-.031 }$ \\[.2cm]$0.965\substack{ +.028/.000 \\ -.021/-.031 }$\end{tabular}& \begin{tabular}{@{}c@{}}$1.004\substack{ +.066/.000 \\ -.046/-.031 }$ \\[.2cm]$0.984\substack{ +.034/.000 \\ -.030/-.031 }$\end{tabular}& \begin{tabular}{@{}c@{}}$0.952\substack{ +.024/.000 \\ -.020/-.019 }$ \\[.2cm]$0.979\substack{ +.015/.000 \\ -.012/-.019 }$\end{tabular}\\[1cm] 
\addlinespace[1ex] 
\addlinespace[1ex] 
\hline 
\hline 
\end{tabular} 
\end{table*} 
\end{turnpage}

\begin{table*}[] 
\centering 
\caption{PTE statistic defined as the fraction of signal+noise
        simulations having $\rl^{BB}$ less than the observed value.
         PTEs do not account for systematic uncertainty.} 
\label{tab:PTE} 
\begin{tabular*}{0.75\textwidth}{@{\extracolsep{\fill}} lcccccccccc} 
\addlinespace[1ex] 
\hline 
\hline 
\addlinespace[1ex] 
\rule{0pt}{2ex} 
  & \mc{LR16}& \mc{LR24}& \mc{LR33}& \mc{LR42}& \mc{LR53}& \mc{LR63N}& \mc{LR63}& \mc{LR63S}& \mc{LR72} \\ 
\addlinespace[1ex] 
\hline 
\addlinespace[1ex] 
  $f_{\rm sky}^{\rm eff}$ [\%]\hfil& 16& 24& 33& 42& 53& 33& 63& 30& 72\\ 
\addlinespace[1ex] 
\hline 
\hline 
\addlinespace[1ex] 
 & \multicolumn{9}{c}{PTE$_{BB}$ \Big( \begin{tabular}{@{}c@{}} HM \\[.05cm] DS \end{tabular} \Big)   } \\\addlinespace[1ex] 
\hline 
\addlinespace[1ex] 
\addlinespace[1ex] 
 \begin{tabular}{@{}c@{}}50--160\\ (no d.b.) \end{tabular} & \begin{tabular}{@{}c@{}}0.020 \\0.000\end{tabular}& \begin{tabular}{@{}c@{}}0.016 \\0.002\end{tabular}& \begin{tabular}{@{}c@{}}0.010 \\0.052\end{tabular}& \begin{tabular}{@{}c@{}}0.004 \\0.002\end{tabular}& \begin{tabular}{@{}c@{}}0.002 \\0.000\end{tabular}& \begin{tabular}{@{}c@{}}0.000 \\0.004\end{tabular}& \begin{tabular}{@{}c@{}}0.002 \\0.000\end{tabular}& \begin{tabular}{@{}c@{}}0.146 \\0.016\end{tabular}& \begin{tabular}{@{}c@{}}0.006 \\0.000\end{tabular}\\[.5cm] 
\addlinespace[1ex] 
 50--160 & \begin{tabular}{@{}c@{}}0.044 \\0.022\end{tabular}& \begin{tabular}{@{}c@{}}0.036 \\0.008\end{tabular}& \begin{tabular}{@{}c@{}}0.038 \\0.190\end{tabular}& \begin{tabular}{@{}c@{}}0.014 \\0.024\end{tabular}& \begin{tabular}{@{}c@{}}0.004 \\0.000\end{tabular}& \begin{tabular}{@{}c@{}}0.006 \\0.044\end{tabular}& \begin{tabular}{@{}c@{}}0.012 \\0.010\end{tabular}& \begin{tabular}{@{}c@{}}0.230 \\0.034\end{tabular}& \begin{tabular}{@{}c@{}}0.032 \\0.014\end{tabular}\\[.5cm] 
\addlinespace[1ex] 
 20--55 & \ldots& \ldots& \ldots& \ldots& \ldots& \begin{tabular}{@{}c@{}}0.584 \\0.938\end{tabular}& \begin{tabular}{@{}c@{}}0.024 \\0.912\end{tabular}& \begin{tabular}{@{}c@{}}0.000 \\0.632\end{tabular}& \begin{tabular}{@{}c@{}}0.018 \\0.894\end{tabular}\\[.5cm] 
\addlinespace[1ex] 
 55--90 & \begin{tabular}{@{}c@{}}0.612 \\0.076\end{tabular}& \begin{tabular}{@{}c@{}}0.566 \\0.164\end{tabular}& \begin{tabular}{@{}c@{}}0.804 \\0.022\end{tabular}& \begin{tabular}{@{}c@{}}0.690 \\0.052\end{tabular}& \begin{tabular}{@{}c@{}}0.668 \\0.148\end{tabular}& \begin{tabular}{@{}c@{}}0.700 \\0.400\end{tabular}& \begin{tabular}{@{}c@{}}0.746 \\0.488\end{tabular}& \begin{tabular}{@{}c@{}}0.640 \\0.396\end{tabular}& \begin{tabular}{@{}c@{}}0.748 \\0.624\end{tabular}\\[.5cm] 
\addlinespace[1ex] 
 90--125 & \begin{tabular}{@{}c@{}}0.066 \\0.002\end{tabular}& \begin{tabular}{@{}c@{}}0.074 \\0.000\end{tabular}& \begin{tabular}{@{}c@{}}0.074 \\0.288\end{tabular}& \begin{tabular}{@{}c@{}}0.022 \\0.076\end{tabular}& \begin{tabular}{@{}c@{}}0.010 \\0.002\end{tabular}& \begin{tabular}{@{}c@{}}0.436 \\0.288\end{tabular}& \begin{tabular}{@{}c@{}}0.086 \\0.012\end{tabular}& \begin{tabular}{@{}c@{}}0.052 \\0.008\end{tabular}& \begin{tabular}{@{}c@{}}0.188 \\0.026\end{tabular}\\[.5cm] 
\addlinespace[1ex] 
 125--160 & \begin{tabular}{@{}c@{}}0.118 \\\ldots\end{tabular}& \begin{tabular}{@{}c@{}}0.061 \\0.499\end{tabular}& \begin{tabular}{@{}c@{}}0.038 \\0.414\end{tabular}& \begin{tabular}{@{}c@{}}0.034 \\0.150\end{tabular}& \begin{tabular}{@{}c@{}}0.024 \\0.084\end{tabular}& \begin{tabular}{@{}c@{}}0.000 \\0.024\end{tabular}& \begin{tabular}{@{}c@{}}0.018 \\0.068\end{tabular}& \begin{tabular}{@{}c@{}}0.560 \\0.290\end{tabular}& \begin{tabular}{@{}c@{}}0.018 \\0.056\end{tabular}\\[.5cm] 
\addlinespace[1ex] 
\hline 
\hline 
\end{tabular*} 
\end{table*}

\end{document}